\documentclass[preprint,nofootinbib,aps,superscriptaddress,eqsecnum]{revtex4-1} 
\pdfoutput=1
\usepackage{extarrows}
\usepackage{graphicx}
\usepackage{amsmath}
\usepackage{amsfonts}
\usepackage{amssymb}
\usepackage{hyperref}
\usepackage{caption}
\usepackage{subcaption}
\usepackage{color}
\allowdisplaybreaks
\def\bea{\begin{eqnarray}}
\def\eea{\end{eqnarray}}
\def\be{\begin{equation}}
\def\ee{\end{equation}}

\begin{document}

	\title{Lepton flavor violation and leptogenesis in discrete flavor symmetric scotogenic model}
	\author{Bichitra Bijay Boruah}
	\email{bijay@tezu.ernet.in}
	\author{Lavina Sarma}
	\email{lavina@tezu.ernet.in}
	\author{Mrinal Kumar Das}
	\email{mkdas@tezu.ernet.in}
	\affiliation{Department of Physics, Tezpur University, Tezpur 784028, India}
		\begin{abstract}
		
	 We have studied the scotogenic model proposed by Ernest Ma, which is an extension of the Standard Model by three singlet right handed neutrinos and a scalar doublet. This model proposes that the light neutrinos acquire a non-zero mass at 1-loop level. In this work, realisation of the scotogenic model is done by using discrete symmetries $A_{4}\times Z_{4}$ in which the non-zero $\theta_{13}$ is produced by assuming a non-degeneracy in the loop factor. Considering different lepton flavor violating(LFV) proceses such as $l_{\alpha}\longrightarrow l_{\beta}\gamma$ and $l_{\alpha}\longrightarrow 3l_{\beta}$, their impact on neutrino phenomenology is studied. We have also analysed $0\nu\beta\beta$ and baryon asymmetry of the Universe(BAU) in this work.
	\end{abstract} 
	\maketitle
	
		\section{Introduction:} Standard model(SM) has been the most successful model of particle physics which is supported by a large amount of experimental evidences. But, SM fails to provide proper explaination  for
	some imporant problems of modern physics, such as absolute neutrino mass\cite{deSalas:2017kay}, baryon asymmetry of universe(BAU)\cite{leptogenesis,Hugle:2018qbw}, dark matter\cite{bertone2005particle,Moore:1999nt} etc. The quest for these unexplained physical problems results in the several SM extensions, aiming at a common explanation for these issues. The scotogenic model proposed by Ernest Ma\cite{Ma:2006km} is one such attractive framework where SM is minimally extended. In this framework, SM is extended by a scalar doublet and three singlet fermions which are charged under $Z_{2}$ symmetry. This inbuilt discrete symmetry forbids the usual tree level generation of neutrino mass. Neutrino mass is generated at 1-loop level within this model. Phenomenology of scotogenic model has been adressed in various literatures\cite{Borah:2017dfn,Borah:2018rca,Merle:2015ica,Lavina}.
	
	 The scotogenic model is probably the simplest TeV scale model that can simultaneously account for neutrino masses and dark matter. In this model, lepton flavor violating processes,
	such as $\mu \rightarrow e\gamma$ and $\mu \rightarrow 3 e$, also takes place at 1-loop, via diagrams analogous to
	those responsible for neutrino masses. Lepton flavor violation(LFV) in the scotogenic model has already been studied\cite{Vicente:2014wga,Takashi}. However, most of the literature primariliy foucses on  $\mu \rightarrow e\gamma$ due to its stringent experimental limit whereas other decays are rarely considered. Here, in this work, we have LFV in details along with other phenomenologies. We know that the baryon asymmetry of universe is one the most important problems which cannot be adressed in SM as it fails to satisfy the Sakharov conditions \cite{Sakharov:1967dj}, which demands baryon number (B) violation, C and CP violation, and departure from thermal equilibrium. We are therefore, intersested to study BAU in the framework of scotogenic model. We can incorporate baryogenesis via leptogenesis \cite{leptogenesis}, where a net leptonic asymmetry is generated first, which further gets converted into baryogenesis through $(B+L)$ violating electroweak sphaleron phase transitions \cite{EWSphaleron}. As discussed in many literatures\cite{Davidson:2002qv,Buchmuller:2002rq}, it is known that there exists a lower bound of about 10TeV for the lightest of the right handed neutrino in the Scotogenic model considering the vanilla leptogenesis scenario\cite{Hugle:2018qbw,Borah:2018rca}. Thus, in our work, a similar kind of leptogenesis takes place. One of the most important significance of BSM frameworks is to study the origin of neutrino mass along with charge lepton mixing and identifying possible symmetries related to this. By incorporating symmetries to a model one can make a model more general and predictive as it will corelate two or more free parameters of the model or make them vanish. Discerete flavor symmetric realisation of sctogenic model is done in a very few work\cite{Adulpravitchai:2009gi,Pramanick:2019qpg}.

	In this work we realised the scotogenic model through $A_{4}\times Z_{4}$ discrete flavor symmetry. The implications of the discrete symmetry can be seen as it contraints the Yukawa couplings of a particular model. Here, we produce a realistic neutrino mixing to do an extensive analysis of lepton flavor violating processes. Considering different lepton flavor violating(LFV) proceses such as $l_{\alpha}\rightarrow l_{\beta}\gamma$ and $l_{\alpha}\rightarrow 3l_{\beta}$, we analysed their impact on the neutrino phenomenology as well. The most stringent bounds on LFV comes from the MEG experiment\cite{TheMEG}. The limit on branching ratio for the decay of $\mu \rightarrow e\gamma$ from this experiment is obtained to be Br($\mu \rightarrow e\gamma$)$< 4.2\times 10^{-13}$. In case of $l_{\alpha}\rightarrow 3 l_{\beta}$ decay contraints comes from SINDRUM experiment\cite{Perrevoort:2018cqi} is set to be $\rm BR(l_{\alpha} \rightarrow 3l_{\beta})<10^{-12}$. Neutrinoless double beta decay($0\nu\beta\beta$) is also studied within the model by the consideration of the constraints from KamLAND-Zen experiment. We have also incorporated BAU within the model and have shown the viable parameter space satisfying the Planck bound.
	
	The rest of the paper is organized as follows: in section\eqref{s1} we introduce the model
	whereas in section\eqref{s2} we realise the scotogenic model using $A_{4}\times Z_{4}$ symmetry. Section\eqref{s3} and \eqref{s4} contains the discussions on LFV processes and leptogenesis respectively.
	Phenomenological analysis of the model is given in section\eqref{s5}. Finally, we summarize our results and draw our
	conclusions in section\eqref{s6} and present additional analytical results in appendices A and B.

		\section{Scotogenic Model}\label{s1}

		Scotogenic model is an extension of the IHDM\cite{Honorez:2010re,LopezHonorez:2006gr,Arhrib:2012ia,Bhattacharya:2019fgs,BorahBR} and the IHDM is nothing but a minimal extension of the SM by a Higgs field which is a doublet under $SU(2)_{L}$ gauge symmetry with hypercharge $Y=1$ and a built-in discrete $Z_{2}$ symmetry \cite{Ma:2006km,Hambye:2009pw,Dolle:2009fn,Honorez:2010re,Gustafsson:2012aj}. The necessity of this modification took place as the inert Higgs doublet model(IHDM) could only accommodate dark matter, whereas it failed in explaining the origin of neutrino masses at a renormalizable level . In this model, three neutral singlet fermions $N_{i}$ with $i=1,2,3$ are added in order to generate neutrino masses and assign them with a discrete $Z_{2}$ symmetry. Here, $N_i$ is odd under $Z_{2}$ symmetry, whereas the SM fields remain $Z_{2}$ even. Symbolic transformation of the particles under $Z_{2}$ symmetry is given by,
		
		\begin{equation}
		N_{i}\longrightarrow -N_{i},~ \eta\longrightarrow -\eta,~ \phi\longrightarrow \phi,~\Psi\longrightarrow \Psi,
		\end{equation}
		where $\eta$ is the inert Higgs doublet, $\phi$ is the SM Higgs doublet and $\Psi$ denotes the SM fermions.
		The new leptonic and scalar particle content can thereafter be represented as follows under the group of symmetries $SU(2) \times U(1)_{Y} \times Z_{2}$:
		\begin{equation*}
		\begin{pmatrix}
		\nu_{\alpha}\\
		l_{\alpha}
		\end{pmatrix}_{L} \sim (2,  -\dfrac{1}{2}, +) ,~ l^{c}_{\alpha} \sim (1,1,+), ~
		\begin{pmatrix}
		\phi^{+}  \\
		\phi^{0} \\
		\end{pmatrix} \sim (2,\frac{1}{2},+), 
		\end{equation*}
		\begin{equation}
		N_{i} \sim (1,1,-), ~
		\begin{pmatrix}
		\eta^{+}\\
		\eta^{0}\\
		\end{pmatrix} \sim (2,1/2, -).
		\end{equation}
		The scalar doublets are written as follows :
		\begin{equation}
		\eta=
		\begin{pmatrix}
		\eta^{\pm}\\
		\frac{1}{\sqrt{2}}(\eta^{0}_{R} + i\eta^{0}_{I}) \\
		\end{pmatrix}, \quad
		\phi=
		\begin{pmatrix}
		\phi^{+}\\
		\frac{1}{\sqrt{2}}(h+i\xi)
		\end{pmatrix}.
		\end{equation} 
			 \hspace{3cm}
		\begin{figure}[h]
			\centering
			\includegraphics[width=0.4\textwidth]{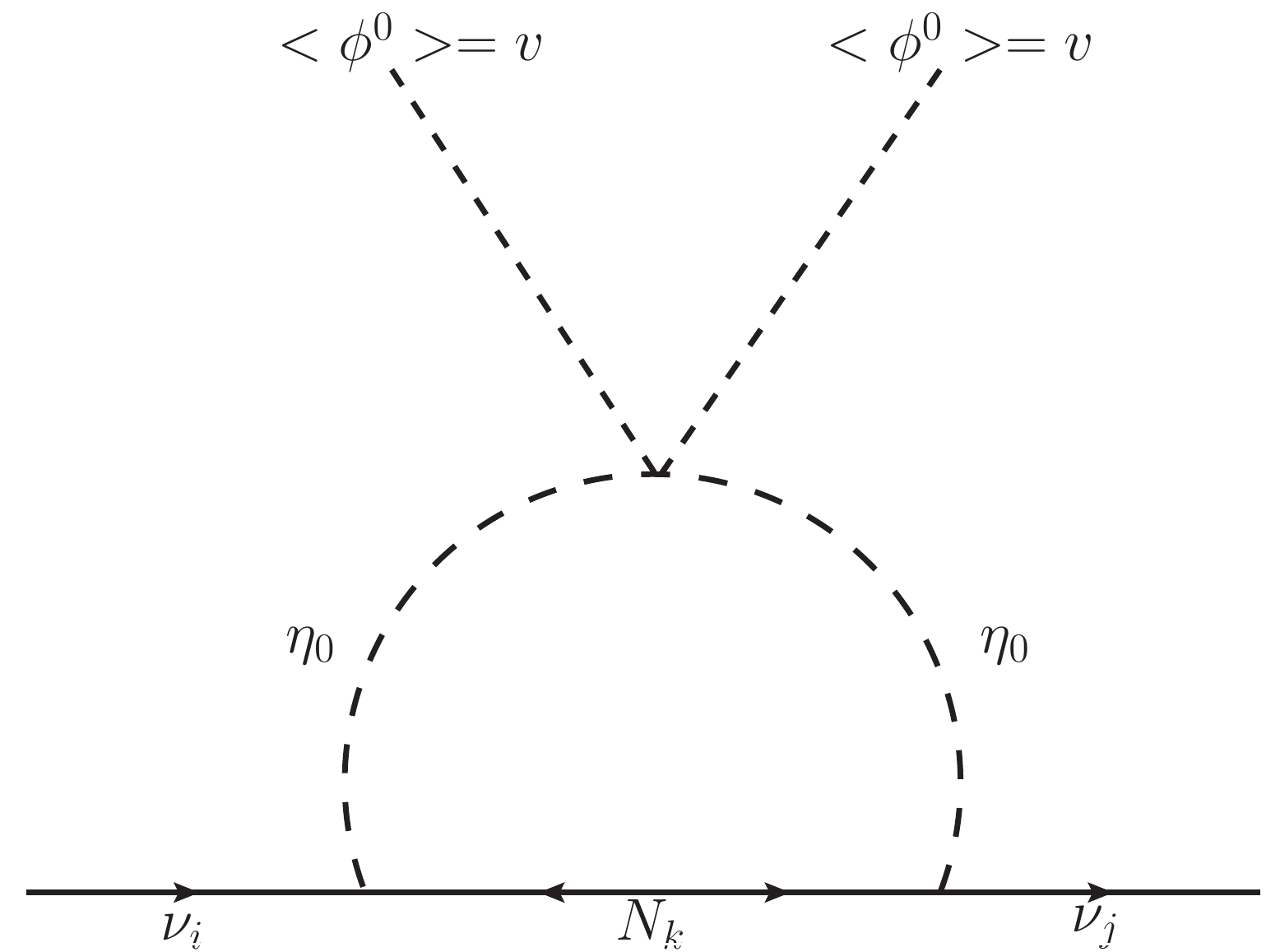}
			\caption{ One- loop contribution of neutrino mass generation with the exchange of right handed neutrino $N_i$ and the scalar $\eta_0$. } \label{fig1}
		\end{figure}\\
		We have no Dirac mass term with $\nu$ and $N_i$, however, the similar Yukawa-like coupling involving $\eta$ is allowed. Nevertheless, the scalar cannot get a VEV. The neutrino mass can be generated through a one-loop mechanism, which is based on the exchange of $\eta$ particle and a heavy neutrino. In figure, we see two Higgs fields $\phi^{0}$ are involved. They will not propagate but will acquire VEV after the EWSB.
		The lagrangian involving the newly added field is :
		\begin{equation}
		\mathcal{L} \supset \frac{1}{2}(M_{N})_{ij}N_{i}N_{j} + Y_{ij}\bar{L}
		\tilde{\eta}N_{j} + h.c 
		\end{equation}
		where, the $1^{st}$ term is the Majorana mass term for the neutrino singlet and the $2^{nd}$ term is the Yukawa interactions of the lepton.
		The new potential on addition of the new inert scalar doublet is:
		\begin{equation}
		\begin{split}
		V_{Scalar} = &m_{1}^{2}\phi^{+}\phi + m_{2}^{2}\eta^{+}\eta + \frac{1}{2}\lambda_{1}(\phi^{+}\phi)^{2} + \frac{1}{2}\lambda_{2}(\eta^{+}\eta)^{2}+ \lambda_{3}(\phi^{+}\phi)(\eta^{+}\eta) \\ &+
		\lambda_{4}(\phi^{+}\eta)(\eta^{+}\phi) 
		+\frac{1}{2}\lambda_{5}[(\phi^{+}\eta)^{2} + h.c.]\label{eqp1}
		\end{split}
		\end{equation}
		All the parameters in Eq. \eqref{eqp1} are real by hermicity of the Lagrangian, except for $\lambda_{5}$. Since, the bilinear term $(\phi^{+}\eta)$ is forbidden by the exact $Z_{2}$ symmetry, therefore one can always choose $\lambda_{5}$ real by rotating the relative phase between $\phi$ and $\eta$. Furthermore, after the spontaneous symmetry breaking like in the SM, we are left with one physical Higgs boson \textit{h} which resembles the SM Higgs boson, as well as four dark scalars: one CP even($\eta^{0}_{R}$), one CP odd($\eta_{I}^{0}$) and a pair of charged ones ($\eta^{\pm}$). The masses of these physical scalars are:
		\begin{equation}
		\begin{split}
		m^{2}_h =& -m^{2}_{1} = 2\lambda_{1}\textit{v}^{2},\\
		m^{2}_{\eta^{\pm}} =& m^{2}_{2}+\lambda_{3}\textit{v}^{2},\\
		m^{2}_{\eta_{R}^{0}} =& m^{2}_{2} + (\lambda_{3}+\lambda_{4}+\lambda_{5})\textit{v}^{2},\\
		m^{2}_{\eta_{I}^{0}} = &m^{2}_{2} + (\lambda_{3}+\lambda_{4}-\lambda_{5})\textit{v}^{2}.
		\end{split}
		\end{equation}
		It is clear from the above equations that all the scalar couplings are written in terms of physical scalar masses and $m_{2}$ , thereby providing six independent parameters of the model to be :
		$\{m_{2},m_{\textit{h}},m_{\eta_{R}^{0}},m_{\eta_{I}^{0}},m_{\eta^{\pm}},\lambda_{2} \}{\large {\tiny }} $.
		Here, $m_{\textit{h}}$ is the mass of SM-Higgs, $m_{\eta_{R}^{0}}$, $m_{\eta_{I}^{0}}$ and $m_{\eta^{\pm}}$ are the masses of CP-even, CP-odd and charged scalars of the inert doublet respectively. In this work, as we have considered the CP-even scalar to be the lightest particle and a probable DM candidate, so we consider $\lambda_{5} < 0$ without any loss of generality. Also, the limit $\lambda_{5}\rightarrow 0$ leads to the mass degeneracy of the neutral components of the inert doublet. Following the 't Hooft scenario\cite{tHooft:1980xss} , the smallness of $\lambda_{5}$ to obtain the lepton asymmetry, which would have been lost if considered to be zero, is acceptably natural.  
		We have a simplified diagram shown in Fig.\eqref{fig3} that can be split further into two diagrams and from which the mass can be easily calculated by considering mechanism after EWSB. 
		 \hspace{3cm}
		\begin{figure}[h]
			\centering
			\includegraphics[width=0.4\textwidth]{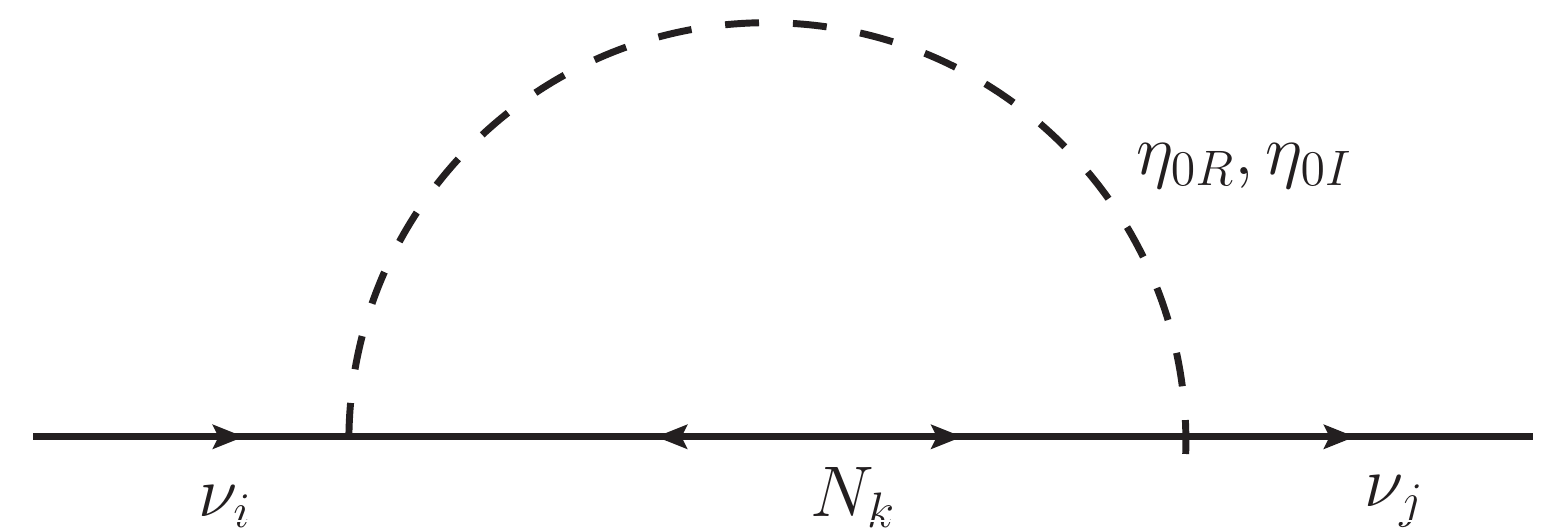}
			\caption{ One-loop diagram with exchange of $\eta_{R}^{0}$ and $\eta_{I}^{0}$. $\nu_i$ and $\nu_j$ representing two different generations of active neutrinos. $N_i$ is the right handed neutrino. } \label{fig3}
		\end{figure}\\
		
		Calculation on the basis of one diagram is sufficient and considered as other would be same except for $\eta_{R}^{0}$ replaced by $\eta_{I}^{0}$. The neutrino mass matrix arising from the radiative mass model is given by :
		\begin{equation}
		\begin{split}
		\textit{M}_{ij}^{\nu}=&\sum_{k} \frac{Y_{ik}Y_{jk}}{16\pi^{2}}M_{\textit{N}_\textit{k}}\left[\frac{m_{\eta_{R}^{0}}^{2}}{m_{\eta_{R}^{0}}^{2}-M_{\textit{N}_\textit{k}}^{2}}\:ln\frac{m_{\eta_{R}^{0}}^{2}}{M_{\textit{N}_\textit{k}}^{2}}-\frac{m_{\eta_{I}^{0}}^{2}}{m_{\eta_{I}^{0}}^{2}-M_{\textit{N}_\textit{k}}}^{2}\:ln\frac{m_{\eta_{I}^{0}}^{2}}{M_{\textit{N}_\textit{k}}}^{2}\right]\\
		\equiv& \sum_{k} \frac{Y_{ik}Y_{jk}}{16\pi^{2}}M_{\textit{N}_\textit{k}}[L_{k}(m^{2}_{\eta_{R}^{0}}) - L_{k}(m^{2}_{\eta_{I}^{0}})],\label{eq7}
		\end{split}
		\end{equation}
		where $M_{k}$ represents the mass eigenvalue of the mass eigenstate $N_{k}$ of the neutral singlet fermion $N_{k}$ in the internal line with indices j=1,2,3 running over the three neutrino generation with three copies of $N_k$ and Y is the Yukawa coupling matrix. The function $L_{k}(m^{2})$ used in Eq. \eqref{eq7} is given by:
		\begin{equation}
		L_{k}(m^{2})= \frac{m^{2}}{m^{2}-M^{2}_{\textit{N}_\textit{k}}}\ln\frac{m^{2}}{M^{2}_{\textit{N}_\textit{k}}}
		\end{equation}

	\section{Flavor Symmetric realisation of Scotogenic model:}\label{s2}
		In this section we have realised scotogenic model using $A_{4}\times Z_{4}$ flavor symmetry. Scotogenic model already has an inbuilt $Z_{2}$ symmetry which provides the explanation of dark sector within this framework\cite{Ma:2006km}. The particle content and respective charge corresponding to the discrete symmetries are given in Table \ref{TAB}. The discrete symmetries, i.e $A_{4}\times Z_{4}$ will impose constraints on the Yukawa coupling matrix, thereby, constraining the model.

	The Lagrangian for the charged lepton sector is given by:
	\begin{equation}
		\mathcal{L}_{l}= \frac{y_{e}}{\Lambda}({l_{L}}\phi \chi_{T})l^{c}_{R_{1}}+ \frac{y_{\mu}}{\Lambda}({l_{L}}\phi \chi_{T})l^{c}_{R_{2}}+ \frac{y_{\tau}}{\Lambda}({l_{L}}\phi \chi_{T})l^{c}_{R_{3}}
	\end{equation}
	where, $\Lambda$ is the cut-off scale of the theory and $y_{e}$,$y_{\mu}$,$y_{\tau}$ are the coupling constants. Terms in the first parenthesis reperents the product of two triplets($l_{L}$ and $\chi_{T}$) under $A_{4}$, each of these terms contracts with $A_{4}$ singlets $1$,$1^{'}$ and $1^{''}$ corresponding to $l_{R1}$, $l_{R2}$ and $l_{R3}$ respectively. When the flavon $\chi_{T}$ gets vaccum expectation value(vev), flavor symmetry will break and we will get the flavor structure for lepton. Finally it sets the charged lepton coupling matrix as the diagonal one, once the flavon vev as well as Higgs vev are inserted. 
	For the Dirac mass term, the effective Lagrangian can be written as:
	
	\begin{equation}
		\mathcal{L}_{D}= \frac{\eta}{\Lambda}[ Y^{'} l_{L_{i}}N_{i} \chi_{S}+ Y^{''}l_{L_{i}}N_{i} \chi].
	\end{equation}
	Again, the additional right handed neutral fermions are represented by the Lagrangian:
	\begin{equation}
		\mathcal{L}_{M_{N}}= M N_{i}N_{j}\label{MR}.
	\end{equation}
	Now, let us consider the vev allignment of the flavons as follows\cite{Adulpravitchai:2009gi}:\\
	$<\chi_{T}>$= $v_{T}$(1,0,0), $<\chi_{S}>$= $v_{S}$(1,1,1), $<\chi>$= u.  
	With the above considerations, the charged leptonic mass matrix is given by:
	\begin{equation}
		M_{l} =\frac{v_{T}<\phi>}{\Lambda} \begin{pmatrix}
			y_{e}  & 0 & 0 \\
			0 & y_{\mu} & 0\\
			0 & 0 & y_{\tau}\\
		\end{pmatrix}
	\end{equation}
	
	also, the Yukawa coupling matrix takes the form:
	\begin{equation}
		Y = \begin{pmatrix}
			2a+b  & -a & -a \\
			-a & 2a & b-a\\
			-a & b-a & 2a\\
		\end{pmatrix}
	\end{equation}
	where, $ a=Y^{'}\frac{v_{S}}{\Lambda}$ and $ b=Y^{''}\frac{u}{\Lambda}$.
	From Eq.\ref{MR} and by taking into account the degenerate mass spectrum of the right handed neutrinos, we obtain the mass matrix of the form:
	\begin{equation}
		M_{N} = \begin{pmatrix}
			M  & 0 & 0 \\
			0 & 0 & M\\
			0 & M & 0\\
		\end{pmatrix}.
	\end{equation}
	Now, in order to transform the mass matrix of RHN to a diagonal one, we go to a different basis with the help of an unitary matrix, U, which is represented by:
	\begin{equation}
		U = \begin{pmatrix}
			0 & 0 & 	1  \\
			0 & 1 & 0  \\
			1 & 0 & 0\\
		\end{pmatrix}.
	\end{equation}
	By this change in basis, we further obtain a change in the Yukawa coupling matrix, which is given by:
	\begin{equation}
		Y = \begin{pmatrix}
			-a & -a & 	2a+b  \\
			b-a & 2a & -a  \\
			2a & b-a & -a\\
		\end{pmatrix}.
	\end{equation}
	\begin{table}
		
		\begin{center}
			\begin{tabular}{|c|c|c|c|c|c|c|c|c|c|c|c|c|}
				
				\hline 
				Field &$l_{L}$ & $l^{c}_{R_{1}}$ & $l^{c}_{R_{2}}$ & $l^{c}_{R_{3}}$ & $N_{i}$ &  $\phi$ & $\eta$ & $\chi_{T}$&$\chi_{S}$&$\chi$ \\ 
				\hline
				$A_{4}$& $3$ & $1$ & $1^{''}$ &$1^{'}$& $3$ & $1$ & $1$ &  $3$ &$3$&$1$\\
				\hline 
				$Z_{4}$& $i$ & $i$ &$i$& $i$ & $-1$ & $1$ &  $1$ &$-1$&$i$&$i$\\
				\hline 
			\end{tabular} 
		\end{center}
		\caption{Fields and their respective transformations
				under the symmetry group of the model.} \textcolor{red}{\label{TAB}}
	\end{table}
	
	Also, it is seen that the charge lepton mass matrix remains diagonal in this new basis:
	\begin{equation}
		M_{l}^{'diag} = U^{\dagger}.M_{l}.U= M_{l}^{diag}.
	\end{equation}

	As already mentioned, the realisation of scotogenic model is done through $A_{4}\times Z_{4}$ flavor symmetry in this study. Within this model, a loop contribution factor $\rm r_{i}$ is adressed via the relation $\rm r_{i}\propto \frac{1}{M_{N_{i}}}$\cite{Soumita,Pramanick:2019qpg}. So, the contribution of right handed neutrino can be given by $\rm diag(r_{1},r_{2},r_{3})$. However, due to the degeneracy in the RHN masses, the loop factor also becomes degenerate.
	Using Eq.\ref{eq7}, the light neutrino mass matrix arising from the loop diagram Fig.\ref{fig1} is of the following from: 
	\begin{equation}
		M_{\nu_{l}} =\begin{pmatrix}
			K_{1}  & K_{2} & K_{2} \\
			K_{2}& K_{3} & K_{4}\\
			K_{2} & K_{4} & K_{3}\\
		\end{pmatrix}\label{eq3.4}
	\end{equation}
	Eq.\ref{eq3.4} results in a $\mu-\tau$ symmetric light neutrino mass matrix. Therefore, we concentrate on generation of realistic neutrino mixing i.e. non zero $\theta_{13}$, which requires deviation from exact $\mu-\tau$ symmetric mass matrix. To break the $\mu-\tau$ symmetry we have to consider the non-degenerate right handed neutrino mass spectrum. Firstly we will take the condition $\rm r_{1}\neq r_{2}= r_{3}=r$ and further  split the degeneracy of $N_{2}$ and $N_{3}$ by a small amount d, i.e  $\rm r_{3}=r_{2}+\rm d$. Now the structure of the light neutrino mass matrix given in eq\eqref{eq3.4} will deviate  by say $M^{0}$ which is proportional to $\rm d$. So the elements of light neutrino mass matrix(M) after considering $\rm r_{1}\neq r_{2}= r_{3}=r$ are given as follows:


	\begin{equation}
		\begin{aligned}
			\rm	m_{11}=\frac{\pi^{2}}{16}\bigg[a^{2}r_{1}\bigg\{\frac{-P\log(r_{1}^{2}P)}{r_{1}^{2}P-1}+\frac{Q\log(r_{1}^{2}Q)}{r_{1}^{2}Q-1}\bigg\}+a^{2}r\bigg\{\frac{-P\log(r^{2}P)}{r^{2}P-1}+\frac{Q\log(r^{2}Q)}{r^{2}Q-1}\bigg\}\\\rm +(2a+b)^2r\bigg\{\frac{-P\log(r^{2}P)}{r^{2}P-1}+\frac{Q\log(r^{2}Q)}{r^{2}Q-1}\bigg\}\bigg]
		\end{aligned}
	\end{equation}
	
	\begin{equation}
		\begin{aligned}
			\rm m_{12}=a\frac{\pi^{2}}{16}\bigg[-(-a+b)r_{1}\bigg\{\frac{-P\log(r{1}^{2}P)}{r_{1}^{2}P-1}+\frac{Q\log(r_{1}^{2}Q)}{r_{1}^{2}Q-1}\bigg\}-2ar\bigg\{\frac{-P\log(r^{2}P)}{r^{2}P-1}+\frac{Q\log(r^{2}Q)}{r^{2}Q-1}\bigg\}\\ \rm
			-(2a+b)r\bigg\{\frac{-P\log(r^{2}P)}{r^{2}P-1}+\frac{Q\log(r^{2}Q)}{r^{2}Q-1}\bigg\}\bigg]
		\end{aligned}
	\end{equation}
	
	\begin{equation}
		\begin{aligned}
			\rm m_{13}=a\frac{\pi^{2}}{16}\bigg[-2ar_{1}\bigg\{\frac{-P\log(r_{1}^{2}P)}{r_{1}^{2}P-1}+\frac{Q\log(r_{1}^{2}Q)}{r_{1}^{2}Q-1}\bigg\}-(-a-b)r\bigg\{\frac{-P\log(r^{2}P)}{r^{2}P-1}+\frac{Q\log(r^{2}Q)}{r^{2}Q-1}\bigg\}\\ \rm
			-(2a+b)r\bigg\{\frac{-P\log(r^{2}P)}{r^{2}P-1}+\frac{Q\log(r^{2}Q)}{r^{2}Q-1}\bigg\}\bigg]
		\end{aligned}
	\end{equation} 
	\begin{equation}
		\begin{aligned}
			\rm m_{22}=\frac{\pi^{2}}{16}\bigg[(a-b)^{2}r_{1}\bigg\{\frac{-P\log(r_{1}^{2}P)}{r_{1}^{2}P-1}+\frac{Q\log(r_{1}^{2}Q)}{r_{1}^{2}Q-1}\bigg\}+4a^{2}r\bigg\{\frac{-P\log(r^{2}P)}{r^{2}P-1}+\frac{Q\log(r^{2}Q)}{r^{2}Q-1}\bigg\}\\\rm
			+a^{2}r\bigg\{\frac{-P\log(r^{2}P)}{r^{2}P-1}+\frac{Q\log(r^{2}Q)}{r^{2}Q-1}\bigg\}\bigg]
		\end{aligned}
	\end{equation}
	
	\begin{equation}
		\begin{aligned}
			\rm	m_{23}=a\frac{\pi^{2}}{16}\bigg[2(-a+b)r_{1}\bigg\{\frac{-P\log(r_{1}^{2}P)}{r_{1}^{2}P-1}+\frac{Q\log(r_{1}^{2}Q)}{r_{1}^{2}Q-1}\bigg\}+2(-a+b)r\bigg\{\frac{-P\log(r^{2}P)}{r^{2}P-1}+\frac{Q\log(r^{2}Q)}{r^{2}Q-1}\bigg\}\\ \rm +ar\bigg\{\frac{-P\log(r^{2}P)}{r^{2}P-1}+\frac{Q\log(r^{2}Q)}{r^{2}Q-1}\bigg\}\bigg]
		\end{aligned}
	\end{equation}  
	\begin{equation}
		\begin{aligned}
			\rm	m_{33}=\frac{\pi^{2}}{16}\bigg[4a^{2}r_{1}\bigg\{\frac{-P\log(r_{1}^{2}P)}{r_{1}^{2}P-1}+\frac{Q\log(r_{1}^{2}Q)}{r_{1}^{2}Q-1}\bigg\}+(a-b)^{2}r\bigg\{\frac{-P\log(r^{2}P)}{r^{2}P-1}+\frac{Q\log(r^{2}Q)}{r^{2}Q-1}\bigg\}\\ \rm +a^{2}r\bigg\{\frac{-P\log(r^{2}P)}{r^{2}P-1}+\frac{Q\log(r^{2}Q)}{r^{2}Q-1}\bigg\}\bigg]
		\end{aligned}
	\end{equation}   
	\vspace{0.5cm}
	
	where, $\rm P=m^{2}+v^{2}(\lambda_{3}+\lambda_{4}-\lambda_{5})$ and $\rm Q=m^{2}+v^{2}(\lambda_{3}+\lambda_{4}+\lambda_{5})$ with v signifying the vev of the SM Higgs and $\lambda_{2}$, $\lambda_{3}$, $\lambda_{4}$, $\lambda_{5}$ are the quartic couplings. 
	
	Now the final light neutrino mass matrix after splitting the degereracy of $N_{2}$ and $N_{3}$ with a small perturbation d can be written as:
	\begin{equation}
		M_{\nu_{l}}=M+M^{0}
	\end{equation}
	where,
	\begin{equation}
		M^{0} =\rm d\begin{pmatrix}
			0 & 0 & x \\
			0& x& 0\\
		x & 0& 0\\
	\end{pmatrix}\label{eq39}
	\end{equation}
	with $x=2\lambda_{5} v^{2}$.
	\vspace{0.2cm}
	

 \section{Lepton flavor violating processes :}\label{s3}
 No experiment so far has observed a flavor violating process involving charged leptons. However, many experiments are currently going on
 to set strong limits on the most relevant LFV
 observables, in order to constraint parameter space involved in many new physics models. In this section we will discuss various lepton flavor violating processes (LFV) such as $l_{\alpha}\rightarrow l_{\beta}\gamma$,$l_{\alpha}\rightarrow 3 l_{\beta}$ and $\mu-e$ conversion in nuclei\cite{LFV,LFV2}. Currently muon decay experiments are most prominent in nature which provides stringent limits for most models. The MEG collaboration\cite{TheMEG} has been
 able to set the impressive bound on muon decay $\rm BR(l_{\alpha} \rightarrow l_{\beta}\gamma)<4.2 \times 10^{-13}$. This is expected to improve as the experiment is upgraded to MEG II. In case of $l_{\alpha}\rightarrow 3 l_{\beta}$ decay, contraints comes from SINDRUM experiment\cite{Perrevoort:2018cqi} to be $\rm BR(l_{\alpha} \rightarrow 3l_{\beta})<10^{-12}$ which is set long ago. The future Mu3e experiment
 announces a sensitivity of $ 10^{-16}$, which would imply a 4 orders of magnitude improvement on the current bound. 
 
 The neutrinoless $\mu-e$ conversion of muonic atom is the most interseting developments regarding the LFV processes\cite{Dinh:2012bp}. There are many experiments which will basically aim for the positive signal. DeeMe\cite{DeeMee}, Mu2e\cite{Mu2e}, COMET\cite{comet} and PRIME\cite{prime} are such experiments primarily focusing on $\mu-e$ conversion of muonic atom. The sensitivity of these experiments will range from $10^{-14}$ to $10^{-18}$. The current limits on $\tau$ observables are less stringent, but will also get improved in the
 near future by the LHC collaboration , as well as by B-factories such as Belle II\cite{Belle2}.
 \begin{figure}
 	\includegraphics[width=0.3\textwidth]{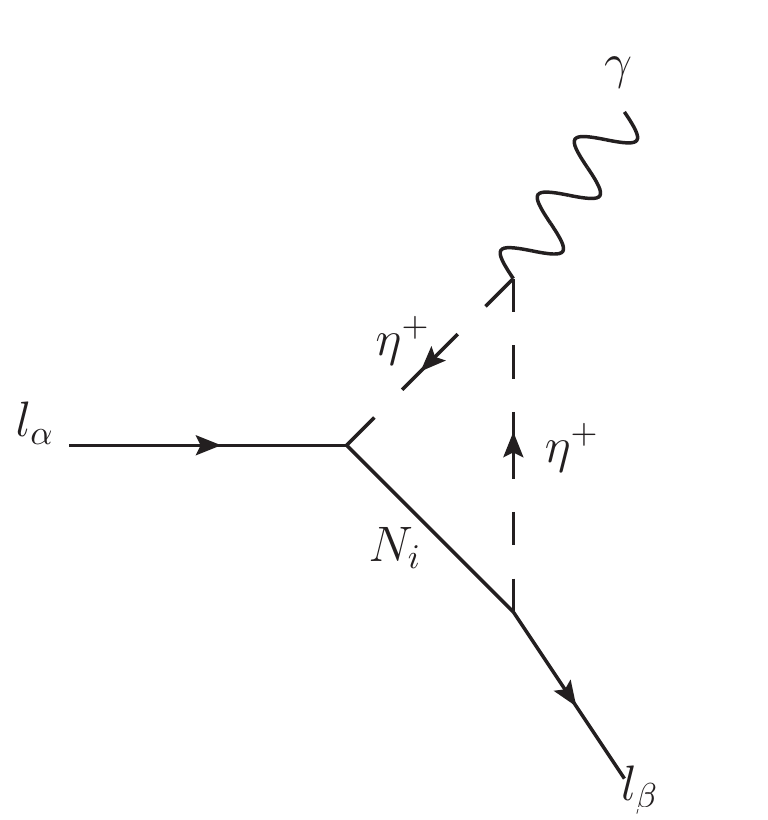}
 	\includegraphics[width=0.3\textwidth]{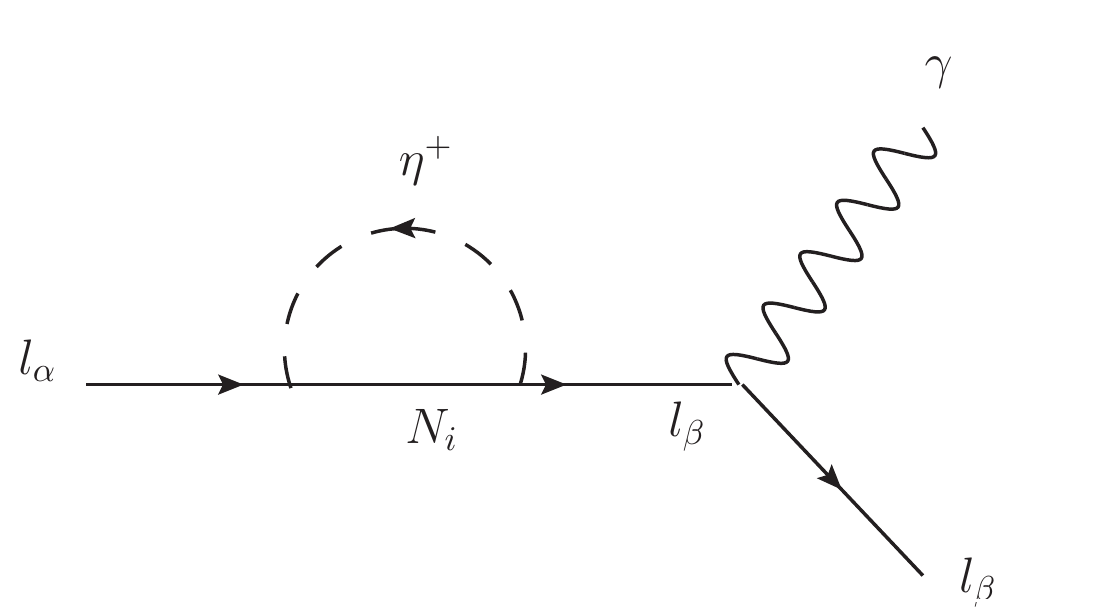}
 	\includegraphics[width=0.3\textwidth]{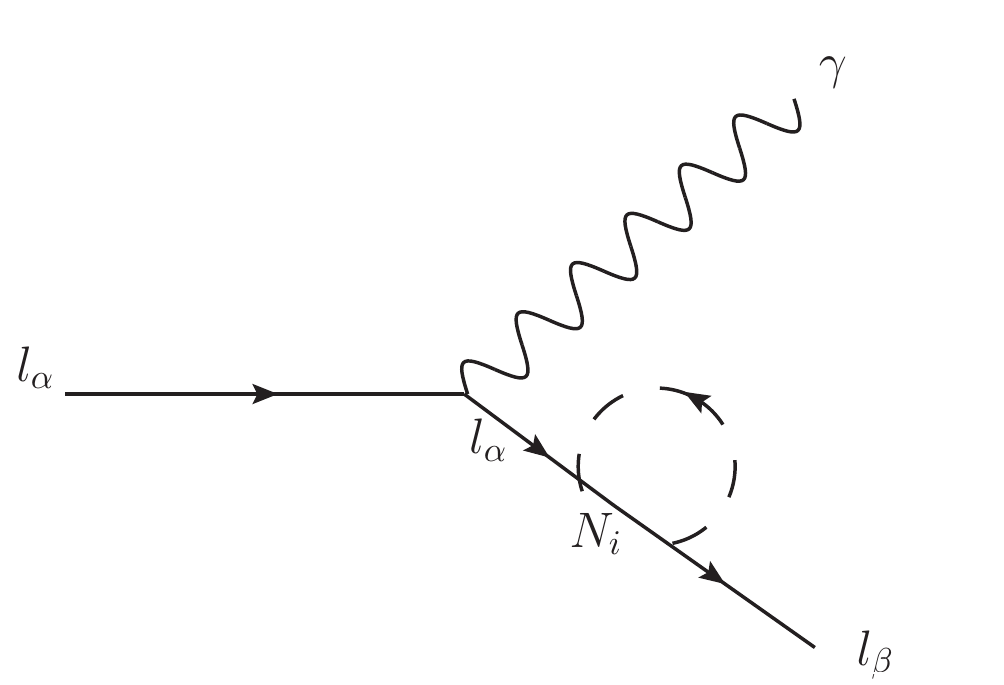}\\
 \caption{ The 1-loop Feynman diagrams giving rise to $l_{\alpha}\longrightarrow l_{\beta}\gamma$.} \label{fig4}
 \end{figure}
 In this part we will discuss the analytical results of branching ratios of different LFV processes such as  $l_{\alpha}\rightarrow l_{\beta}\gamma$,$l_{\alpha}\rightarrow 3 l_{\beta}$ and $\mu-e$ conversion in nuclei within the framework of scotogenic model.
 
 In case of radiative lepton decay, the branching ratio of $l_{\alpha}\rightarrow l_{\beta}\gamma$ is given by\cite{Takashi}-
 
 \begin{equation}
\rm BR(l_{\alpha}\rightarrow l_{\beta}\gamma)=\frac{3(4\pi^{3})\alpha_{em}}{4G_{F}^{2}}|A_{D}|^{2} BR(l_{\alpha} \rightarrow l_{\beta}\nu_{\alpha}\bar{\nu{\beta}}).
 \end{equation}
 Here, $\rm G_{F}$ is the Fermi constant and $\rm \alpha_{em}=\frac{e^{2}}{4\pi}$ is the electromagnetic fine structure constant, with e the electromagnetic coupling. $\rm A_{D}$ is the dipole form factor which is given by-
 
 \begin{equation}
 A_{D}=\sum_{i=1}^{3} \frac{Y_{i\beta}^{*} Y_{i\alpha}}{2(4\pi)^{2}} \frac{1}{m_{\eta^{+}}^{2}} F_{2}(\rho_{i})
 \end{equation}
 
 where the parameter $\rho_{i}$ is defined as $\rho_{i}=\frac{M_{N_{i}}^{2}}{m_{\eta^{+}}^{2}}$ and the loop function $F_{2}(x)$ is given in appendix.
 
 For three body decay process like $l_{\alpha}\rightarrow 3 l_{\beta}$, the branching ratio is given by-
 \begin{equation}
 \begin{aligned}
 \rm BR(l_{\alpha}\rightarrow 3 l_{\beta})=\frac{3(4\pi^{2})\alpha_{em}^{2}}{8G_{F}^{2}} \bigg[|A_{ND}|^{2}+|A_{D}|^{2}\bigg(\frac{16}{3}\log\bigg(\frac{m_{\alpha}}{m_{\beta}}\bigg)-\frac{22}{3}\bigg)\\ 
\rm +\frac{1}{6}|B|^{2}+\bigg(-2A_{ND} A_{D}^{*}+\frac{1}{3}A_{ND} B^{*}-\frac{2}{3}A_{D}B^{*}+h.c\bigg)\bigg] \\ \rm 
\times BR(l_{\alpha} \rightarrow l_{\beta}\nu_{\alpha}\bar{\nu{\beta}}).
 \end{aligned}
 \end{equation}

 Here, we have kept $m_{\beta}<< m_{\alpha}$ only in the logarithmic term, where it avoids the appearance
 of an infrared divergence. The form factor $A_{D}$ is generated by dipole photon penguins and
 is given in equation 17. Regarding the other form factors $A_{ND}$ is  given by-

 \begin{equation}
A_{ND}=\sum_{i=1}^{3} \frac{Y_{i\beta}^{*} Y_{i\alpha}}{6(4\pi)^{2}} \frac{1}{m_{\eta^{+}}^{2}} G_{2}(\rho_{i}).
\end{equation}

$\rm A_{ND}$ is generated by non-dipole photon penguins, whereas B, induced by box diagrams is given by-
\begin{equation}
\rm e^{2}B=\frac{1}{(4\pi)^{2}m_{\eta^{+}}^{2}}\sum_{i,j=1}^{3} \bigg[\frac{1}{2}D_{1}(\rho_{i},\rho_{j})Y_{j\beta}^{*}Y_{j\beta}Y_{i\beta}^{*}Y_{i\alpha}+\sqrt{\rho_{i}\rho_{j}}D_{2}(\rho_{i},\rho_{j})Y_{j\beta}^{*}Y_{j\beta}^{*}Y_{i\beta}Y_{i\alpha}\bigg].
\end{equation}

The loop functions $\rm G_{2}(x)$, $\rm D_{1}(x,y)$ and $\rm D_{2}(x,y)$ are defined in appendix. Here, e Z-boson penguin contributions are negligible, since in this model they
are suppressed by charged lepton masses. Similarly, Higgs-penguin contribution is also supressed which are not taken into consideration.
 \begin{figure}
	\includegraphics[width=0.3\textwidth]{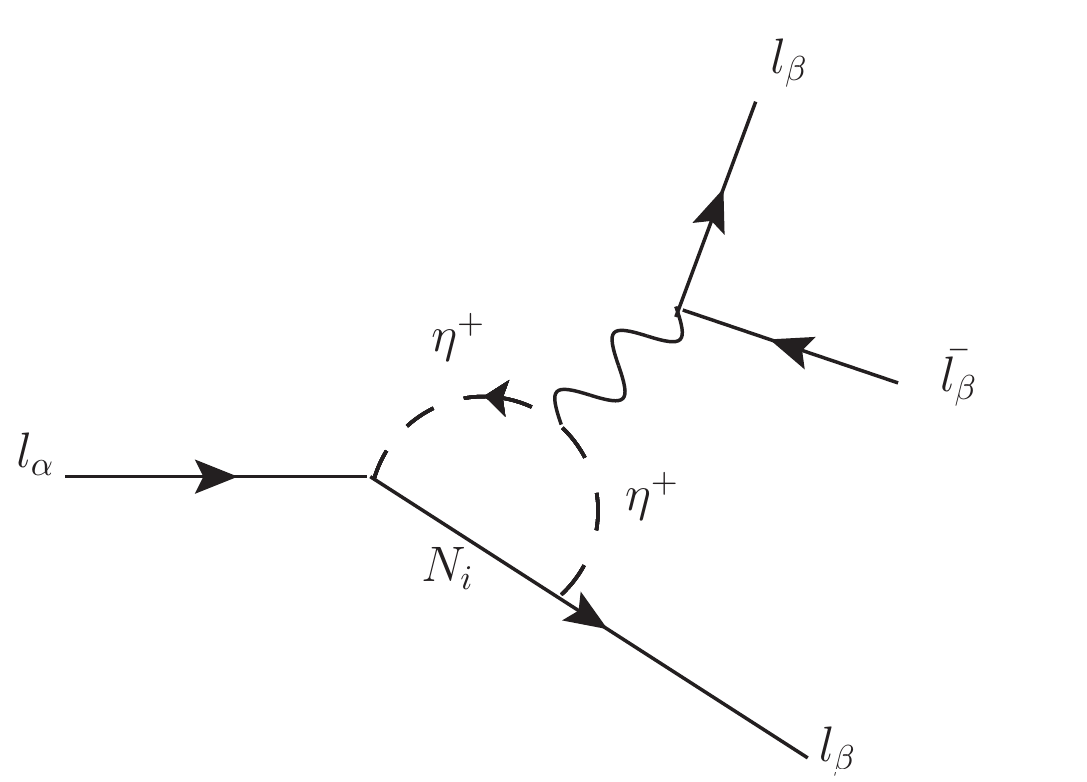}
	\includegraphics[width=0.3\textwidth]{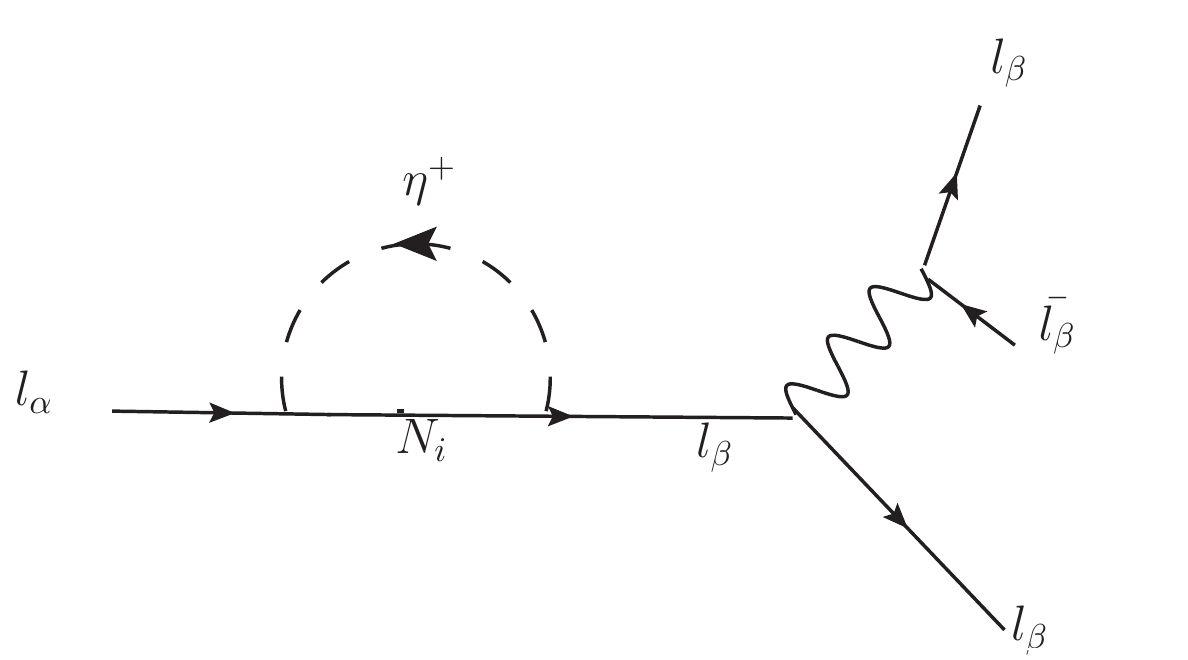}
	\includegraphics[width=0.3\textwidth]{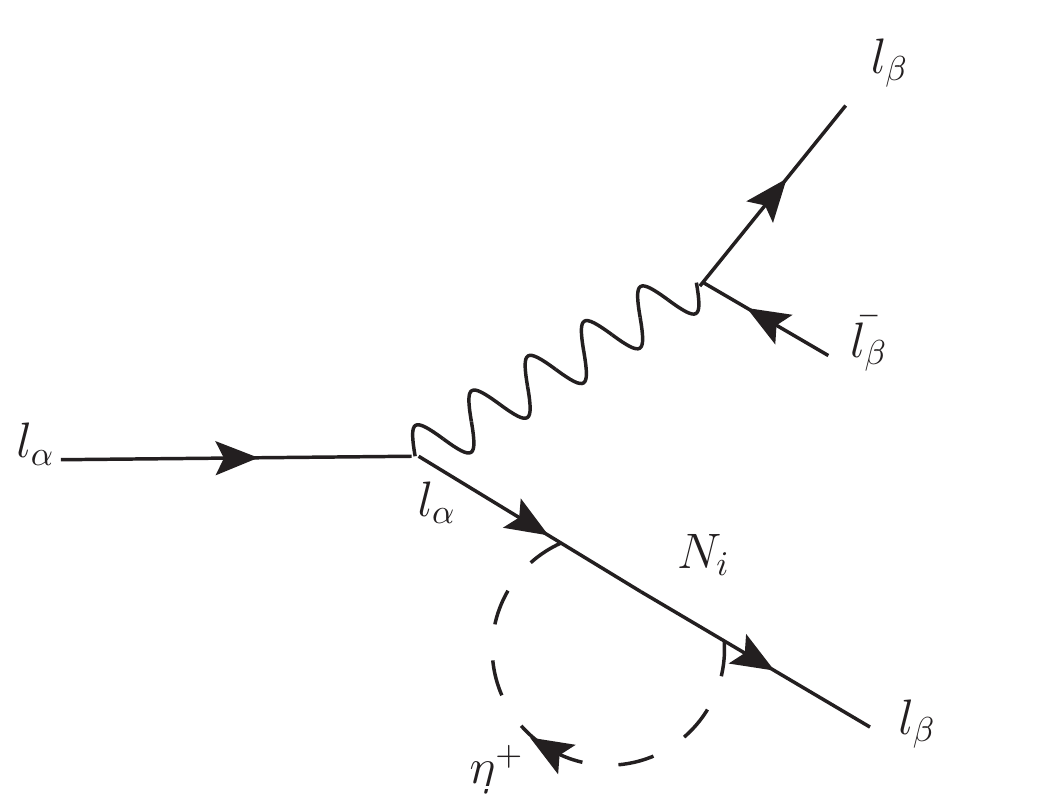}\\
\caption{ The penguin contributions to $l_{\alpha}\longrightarrow l_{\beta}\gamma$, where the wavy lines depicts either a Z-boson or a photon.} \label{fig5}
\end{figure}
Next we consider the case of $\mu-e$ conversion in nuclei. The conversion rate, normalized
to the muon capture rate, can be expressed as -
\begin{equation}
\begin{aligned}
\rm CR(\mu-e,Nucleus)=\frac{p_{e}E_{e}m_{\mu}^{3}G_{F}^{2}\alpha_{em}^{3}Z_{eff}^{4}F_{p}^{2}}{8\pi^{2}Z \Gamma_{capt}}\times \bigg[|(Z+N)(g_{LV}^{(0)}+g_{LS}^{(0)})+(Z-N)(g_{LV}^{(1)}+g_{LS}^{(1)})|^{2}\\ \rm +|(Z+N)(g_{RV}^{(0)}+g_{RS}^{(0)})+(Z-N)(g_{RV}^{(1)}+g_{RS}^{(1)})|^{2}\bigg] 
\end{aligned}
\end{equation}

here, number of photon and neutron is given by Z and N. $\rm Z_{eff}$ is the effective
atomic charge, $F_{p}$ denotes the nuclear matrix element and total muon capture rate is denoted by $\rm \Gamma_{capt}$. The values of these parameter are different for different nucleus under consideration. Also, $\rm p_{e}$ and $E_{e}$ are the momentum and energy of electron. The expression for $g_{XK}^{(0)}$ and $g_{XK}^{(1)}$ (X=L,R and K=S,V) present in the above equation is given by-
\begin{equation}
g_{XK}^{(0)}=\frac{1}{2} \sum_{q=u,d,s}\bigg(g_{XK}(q) G_{K}^{q,p}+g_{XK}(q) G_{K}^{q,n}\bigg)
\end{equation}
\begin{equation}
g_{XK}^{(1)}=\frac{1}{2} \sum_{q=u,d,s}\bigg(g_{XK}(q) G_{K}^{q,p}-g_{XK}(q) G_{K}^{q,n}\bigg).
\end{equation}

Numerical values of $G_{K}$ are given in various literatures. Again the effective couplings $g_{XK}(q)$ in scotogenic model has many contributions, which are given below-
\begin{equation}
g_{LV}(q) \approx g_{LV}^{\gamma}(q) 
\end{equation} 
\begin{equation}
g_{RV}(q) = g_{LV}(q)|_{L\leftrightarrow R}
\end{equation}
\begin{equation}
g_{LS}(q) \approx 0
\end{equation}
\begin{equation}
g_{RS}(q) \approx 0
\end{equation}

where $\rm g_{LV}^{\gamma}(q)$ stands for the contribution due to photon penguins. Because of the inbuilt $Z_{2}$ symmetry in  scotogenic model, there is no box contribution to $\mu-e$ conversion of nuclei. This additional symmetry forbids the coupling between scalars($\eta^{+}$ and $\eta^{-}$) and the quark sector. Regarding the Z-boson penguins contributions, they turn out to be suppressed by charged lepton masses. So the effective coupling can be written as-
\begin{equation}
g_{LV}^{\gamma}(q) =\frac{\sqrt{2}}{G_{F}} e^{2} Q_{p} (A_{ND}-A_{D}).
\end{equation}

The form factors $A_{ND}$ and $A_D$ have been already defined. Furthermore, $Q_p$ is the electric charge of the corresponding quark.

\section{Leptogenesis}\label{s4}
We study baryogenesis in the Scotogenic model realised by $A_{4} \times Z_{4}$ symmetry. We can produce observed baryogenesis via the mechanism of leptogenesis\cite{leptogenesis} in our model. However, the leptogenesis process must occur by the out of equilibrium decay of the RHN, in our case $N_{1}$. As discussed in many literatures\cite{Davidson:2002qv,Buchmuller:2002rq}, we now know that there exists a lower bound of about 10TeV for the lightest of the RHNs($M_{N_{1}}$) in the Scotogenic model considering the vanilla leptogenesis scenario\cite{Hugle:2018qbw,Borah:2018rca}. For a heirarchical mass of RHN, i.e $M_{N_{1}} << M_{N_{2}}, M_{N_{3}}$, the leptogenesis produced by the decay of $N_{2}$ and $N_{3}$ are supressed due to the strong washout effects produced by $N_{1}$ or $N_{2}$ and $N_{3}$ mediated interactions\cite{Borah:2018rca}. Thereby, the lepton asymmetry is produced only by the virtue of $N_{1}$ decay and this is further converted into the baryon asymmetry of the Universe(BAU) by the electro-weak sphaleron phase transitions\cite{Dine:2003ax}. Now for the generation of BAU, we solve the simultaneous Boltzmann equations for $N_{1}$ decay and formation of $N_{B-L}$. The B-L calculation depends on the comparison between the decay rates for $N_{1}\rightarrow l\eta, \bar{l}\eta^{*}$ processes and the Hubble parameter, which causes a certain impact on the asymmetry as well as on the CP-asymmetry parameter $\epsilon_{1}$. In the calculation of leptogenesis, one important quantity that differentiates between weak and strong washout regime is the decay parameter. It is expressed as:
\begin{equation}
K_{1}= \frac{\Gamma_{1}}{H(z=1)},
\end{equation}
where, $\Gamma_{2}$ gives us the total $N_{2}$ decay width, $H$ is the Hubble parameter and $z= \frac{M_{N_{1}}}{T}$ with $T$ being the temperature of the photon bath. We can express $H$ in terms of $T$ and the corresponding equation is given by:
\begin{equation}\label{eq:1}
H = \sqrt\frac{8\pi^{3}g_{*}}{90}\dfrac{T^{2}}{M_{Pl}}.
\end{equation} 
In Eq.\eqref{eq:1}, $g_{*}$ stands for the effective number of relativistic degrees of freedom and $M_{Pl}\simeq 1.22\times 10^{19}$ GeV is the Planck mass. We have introduced a perturbation ($\xi$) in our work to generate non-zero $\theta_{13}$ by breaking the $\mu-\tau$ symmetry. Thus we obtain a non-degeneracy in the RHN masses as discussed in the above section\ref{s2}. The mass of the lightest RHN is fixed in the range $M_{N_{1}}= 10^{4}-10^{5}$ GeV and that of $N_{2}$ is $M_{N_{2}}= 10^{8}-10^{9}$ GeV. The range of $M_{N_{3}}$ will be evaluated considering the relation $M_{N_{3}}$= $M_{N_{2}}+d$.  Now, by this choice of RHN masses along with $m_{\eta^{0}_{R}}= 400-800$ GeV and most significantly the lightest active neutrino mass $m_{1}= 10^{-13}-10^{-12}$ eV, we fall on the weak washout regime. 
The Yukawa couplings obtained by solving the model parameters are incorporated in the decay rate equation for $N_{1}$ which is given by,
\begin{equation}
\Gamma_{1}= \frac{M_{N_{1}}}{8\pi}(Y^{\dagger}Y)_{11}\left[1- \Big(\frac{m_{\eta^{0}_{R}}}{M_{N_{1}}}\Big)^{2}\right]^{2}= \frac{M_{N_{1}}}{8\pi}(Y^{\dagger}Y)_{11}(1-\eta_{1})^{2}
\end{equation}
Again for the decays $N_{1}\rightarrow l\eta, \bar{l}\eta^{*}$, the CP asymmetry parameter $\epsilon_{1}$ is given by, 
\begin{equation}
\epsilon_{1}= \frac{1}{8\pi(Y^{\dagger}Y)_{11}}\sum_{j\ne 1}Im[(Y^{\dagger}Y)^{2}]_{1j}\left[ f(r_{j1},\eta_{1})- \frac{\sqrt{r_{j1}}}{r_{j1}-1}(1-\eta_{1})^{2}\right],
\end{equation}
where,
\begin{equation}
f(r_{j1},\eta_{1})= \sqrt{r_{j1}}\left[1+ \frac{(1-2\eta_{1}+r_{j1})}{(1-\eta_{1})^{2}}  ln(\frac{r_{j1}-\eta_{1}^{2}}{1-2\eta_{1}+r_{j1}})\right],
\end{equation}
and
$r_{j1}= \big(\frac{M_{N_{j}}}{M_{N_{1}}}\big)^{2}$, $\eta_{1}\equiv \big(\frac{m_{\eta^{0}_{R}}}{M_{N_{1}}}\big)^{2}$.\\
The Boltzmann equations for the number densities of $N_{1}$ and $N_{B-L}$, given by \cite{Davidson:2002qv},
\begin{equation}\label{eq:3}
\frac{dn_{N_{1}}}{dz}= -D_{1}(n_{N_{1}} - n_{N_{1}}^{eq}),
\end{equation}
\begin{equation}\label{eq:4}
\frac{dn_{B-L}}{dz}= -\epsilon_{1}D_{1}(n_{N_{1}} - n_{N_{1}}^{eq})- W_{1}n_{B-L},
\end{equation}
respectively. $n_{N_{1}}^{eq}= \frac{z^{2}}{2}K_{1}(z)$ is the equilibrium number density of $N_{1}$, where $K_{i}(z)$ is the modified  Bessel function of $i^{th}$ type and
\begin{equation}
D_{1}\equiv \frac{\Gamma_{1}}{Hz} = K_{N_{1}}z\frac{K_{1}(z)}{K_{2}(z)}
\end{equation} 
gives the measure of the total decay rate with respect to the Hubble rate, and $ W_{1}= \frac{\Gamma_{W}}{Hz}$ is the total washout rate. Again,  $W_{1}= W_{1D}+W_{\Delta L=2}$, i.e the total washout term is the sum of the washout due to inverse decays $l\eta,\bar{l}\eta^{*}\rightarrow N_{1}$ ($W_{1D}= \frac{1}{4}K_{N_{1}}z^{3}K_{1}(z)$) and the washout due to the $\Delta L= 2$ scatterings $l\eta \leftrightarrow \bar{l}\eta^{*},ll \leftrightarrow \eta^{*}\eta^{*}$ which is given by,
\begin{equation}
W_{\Delta L=2} \simeq \dfrac{18\sqrt{10}M_{Pl}}{\pi^{4}g_{l}\sqrt{g_{*}}z^{2}v^{4}}(\frac{2\pi^{2}}{\lambda_{5}})^{2}M_{N_{1}}\bar{m_{\varsigma}}^{2}.
\end{equation}
Here, $g_{l}$ is the internal degrees of freedom for the SM leptons, and $\bar{m_{\varsigma}}$ is the effective neutrino mass parameter, defined by:
\begin{equation}
\bar{m_{\varsigma}}^{2} \simeq 4\varsigma_{1}^{2}m_{1}^{2} + \varsigma_{2}m^{2_{2}} +\varsigma_{3}^{2}m_{3}^{2},
\end{equation}
where $m_{i}'s$ is the light neutrino mass eigenvalues and $ \varsigma_{k}$ is as defined as:
\begin{equation}
\varsigma_{k} = \Big(\frac{M^{2}_{\textit{N}_\textit{k}}}{8(m_{\eta_{R}^{0}}^{2}-m_{\eta_{I}^{0}}^{2})}[L_{k}(m^{2}_{\eta_{R}^{0}}) - L_{k}(m^{2}_{\eta_{I}^{0}})]\Big)^{-1}	
\end{equation} 
The final B-L asymmetry $n_{B-L}^{f}$ is evaluated by numerically calculating Eq.\eqref{eq:3} and Eq.\eqref{eq:4} before the sphaleron freeze-out. This is converted into the baryon-to-photon ratio given by:
\begin{equation}\label{eq:5}
n_{B}= \frac{3}{4}\frac{g_{*}^{0}}{g_{*}}a_{sph}n_{B-L}^{f}\simeq 9.2\times 10^{-3}n_{B-L}^{f},
\end{equation} 
In Eq.\eqref{eq:5}, $g_{*}= 110.75$ is the effective relativistic degrees of freedom at the time when final lepton asymmetry was produced, $g_{*}^{0}= \frac{43}{11}$ is the effective degrees of freedom at the recombination epoch and $a_{sph}=\frac{8}{23}$ is the sphaleron conversion factor taking two Higgs doublet into consideration. The Planck limit 2018 gives a bound on the observed BAU($n_{B}^{obs}$) to be $(6.04\pm0.08)\times 10^{-10}$\cite{Aghanim:2018eyx}. Therefore, in our work we have chosen the free parameters appropriately so as to generate the observed BAU. The values of the quartic coupling, $\lambda_{5}$ is taken in range $10^{-8}- 10^{-4}$ for successful generation of leptogenesis as well as to have significant results for LFV.

\section{Numerical Analysis}\label{s5} 

In our work, we do a random scan for the free parameters of our model given by:\\
\begin{equation}
\rm M_{N_{1}}, \rm M_{N_{2}}, \eta_{R}^{0},\lambda_{5}
\end{equation}
The values of the above mentioned parameters for which we study the impact on neutrino mass, LFV and BAU are given in the Table\ref{TAB1} as follows:\\
\begin{table}[h]
	\begin{center}
		\begin{tabular}{|c|c|}
			
			\hline 
			Parameter & Parameter space \\ 
			\hline
			$\rm M_{N_{1}}$ & $10^{4}$ GeV - $10^{5}$ GeV \\
			\hline 
			$\rm M_{N_{2}}$ & $10^{8}$ GeV - $10^{9}$ GeV\\
			\hline
			$\eta_{R}^{0}$ & $400$ GeV- $800$ GeV\\
			\hline
			$\lambda_{5}$ & $10^{-8}$- $10^{-4}$\\
			\hline   	
		\end{tabular} 
	\end{center}
	\caption{Free parameters of the model and their respective parameter space.} \label{TAB1}
\end{table}\\
\begin{figure}
	\begin{center}
	\includegraphics[width=0.4\textwidth]{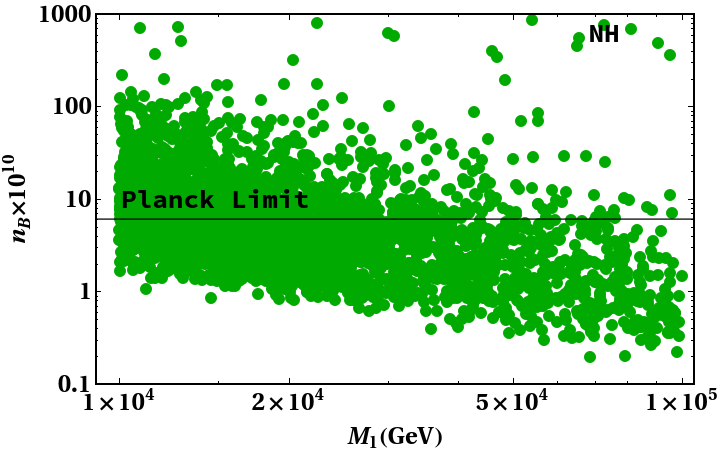}
	\includegraphics[width=0.4\textwidth]{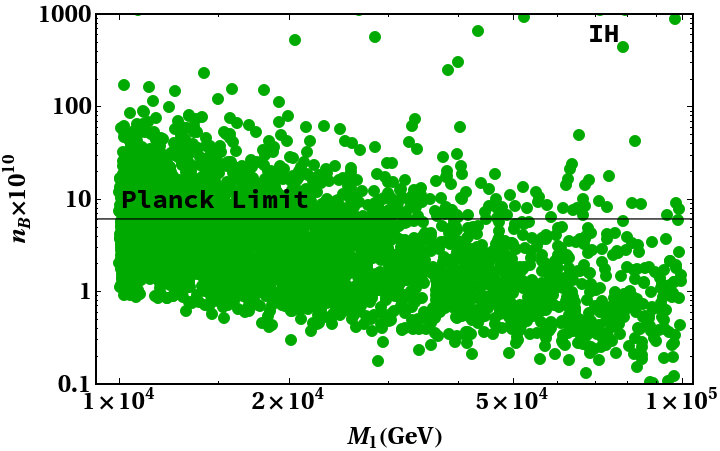}\\
	\includegraphics[width=0.4\textwidth]{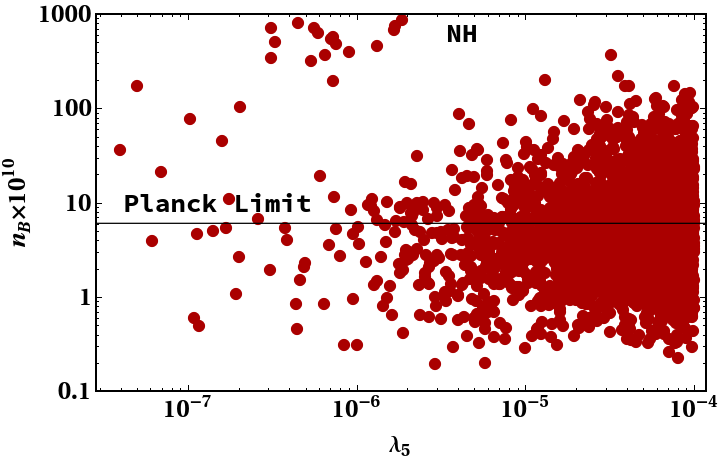}
	\includegraphics[width=0.4\textwidth]{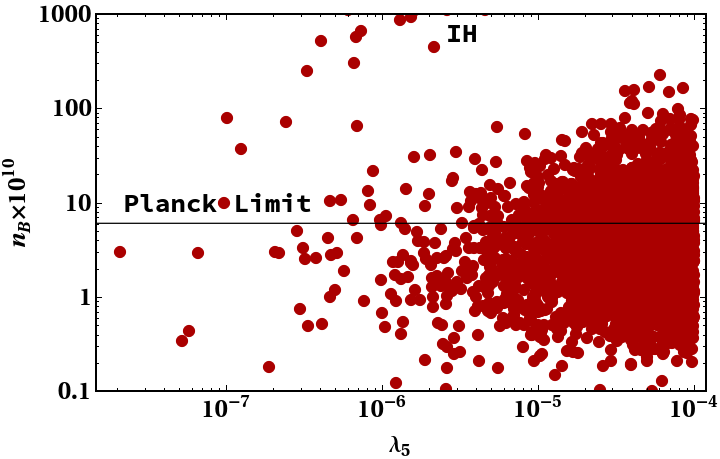}\\
	\includegraphics[width=0.4\textwidth]{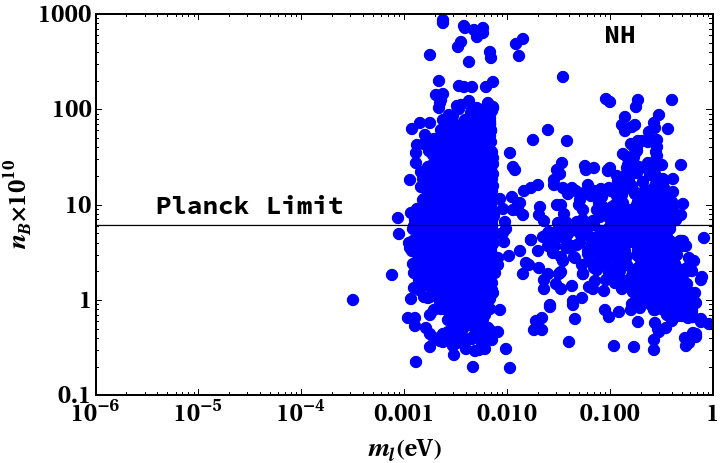}
	\includegraphics[width=0.4\textwidth]{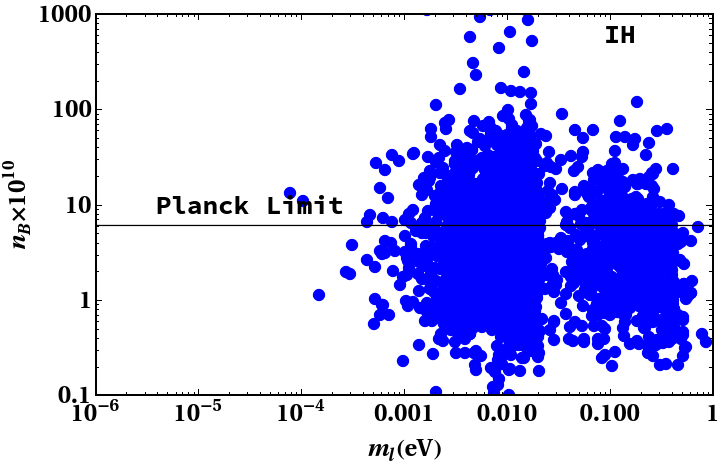}\\
	\includegraphics[width=0.4\textwidth]{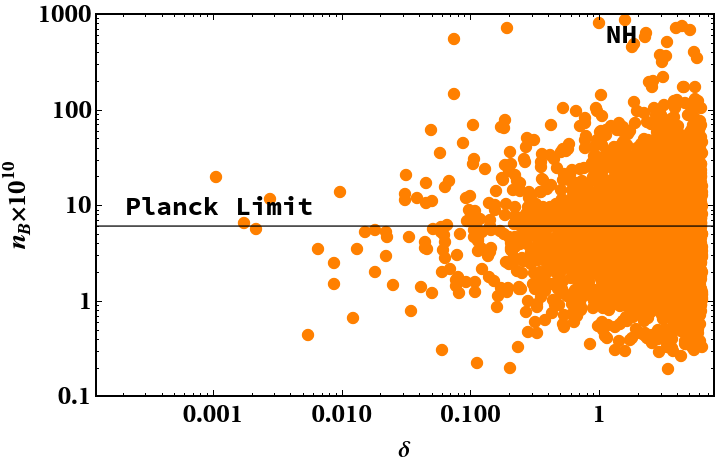}
	\includegraphics[width=0.4\textwidth]{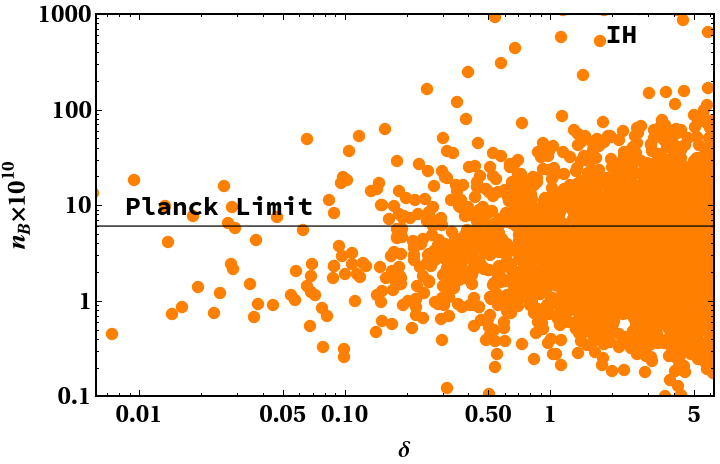}\\

	\caption{Plots in the first-row shows baryon asymmetry as a function of RHN ($M_{N_{1}}$), the second-row shows baryon asymmetry as a function of the quartic coupling ($\lambda_5$), in third-row baryon asymmetry as a function of lightest neutrino mass eigenvalue($m_{l}$) is depicted and in the fourth row a variation between leptonic Dirac CP phase($\delta$) and baryon asymmetry is shown respectively. The black horizontal line gives the current Planck limit for BAU. }\label{bau1}
	\end{center}
\end{figure}
\begin{figure}[h]
	\begin{center}
		\includegraphics[width=0.4\textwidth]{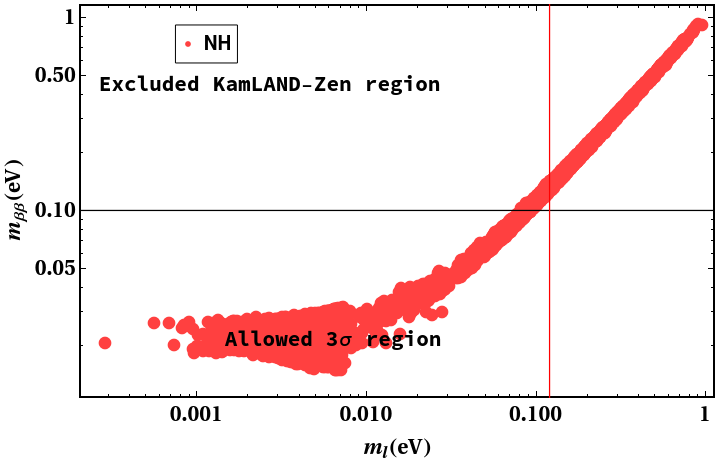}
		\includegraphics[width=0.4\textwidth]{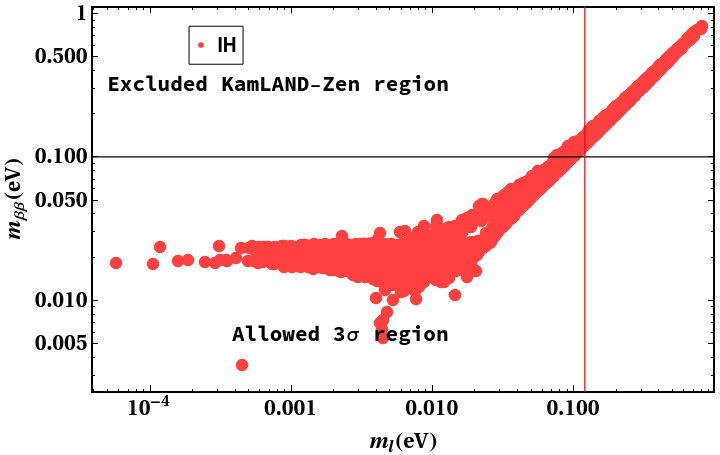}
		\caption{ Effective mass as a function of lightest neutrino mass eigenvalue($m_{l}$) for NH/IH. The horizontal(black) line is the upper limit for the effective mass ($m_{\beta\beta}(eV)\sim0.1(eV)$) of light neutrinos obtained from KamLAND-Zen experiment and the vertical red line depicts the Planck limit for the sum of the light neutrino masses.}\label{eff1}
		
	\end{center}
\end{figure}

\begin{figure}[h]
	\begin{center}
		\includegraphics[width=0.4\textwidth]{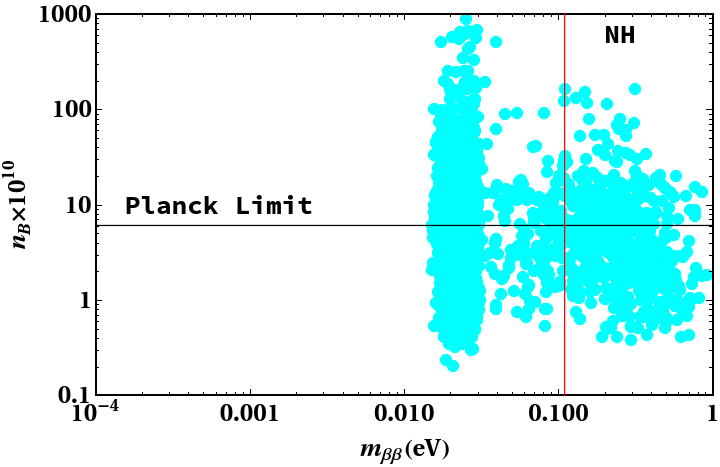}
		\includegraphics[width=0.4\textwidth]{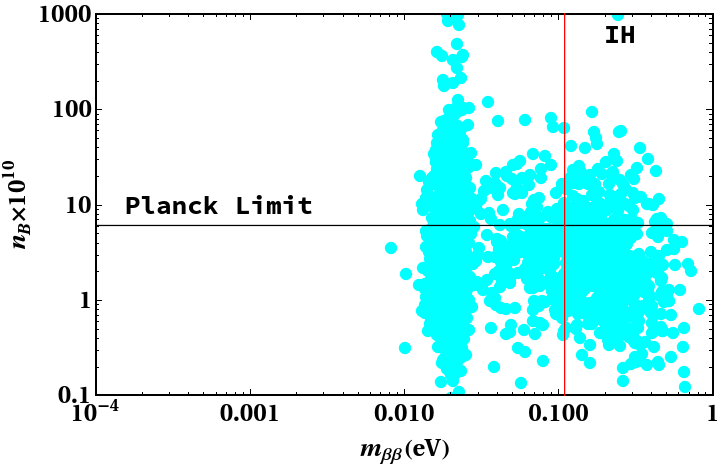}
		\caption{ Baryon asymmetry of the Universe as a function of effective mass of active neutrinos($m_{\beta\beta}$) for NH/IH. The horizontal(black) line is Planck limit on BAU and the vertical red line depicts the upper bound on the effective mass ($m_{\beta\beta}(eV)\sim0.1(eV)$) of light neutrinos obtained from KamLAND-Zen experiment.}\label{eff2}
		
	\end{center}
\end{figure}

\begin{figure}[h]
	\begin{center}
		\includegraphics[width=0.4\textwidth]{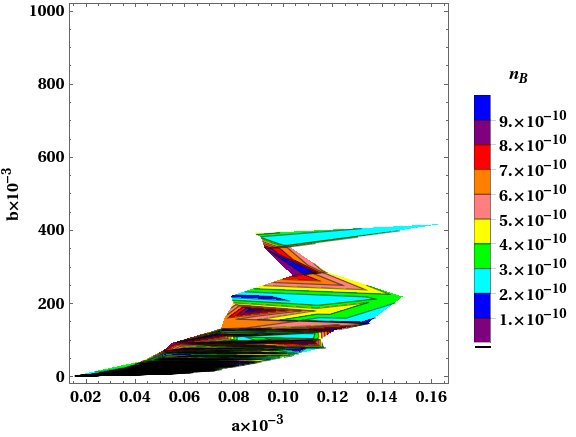}
		\includegraphics[width=0.4\textwidth]{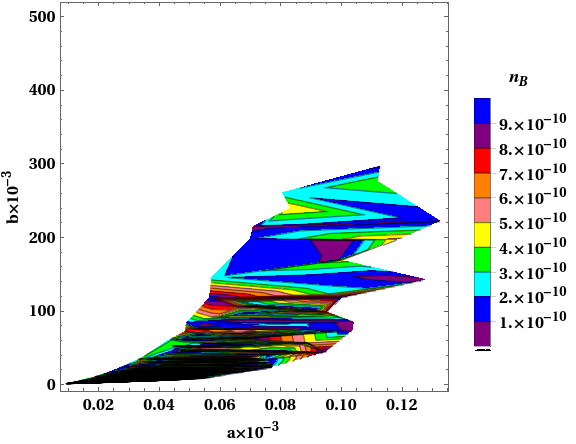}\\
		\caption{Contour plots relating the model parameter a and b to the baryon asymmetry of the Universe. The left panel is for NH and right panel for IH.}\label{cont1}
	\end{center}
\end{figure}

\begin{figure}[h]
	\begin{center}
		\includegraphics[width=0.4\textwidth]{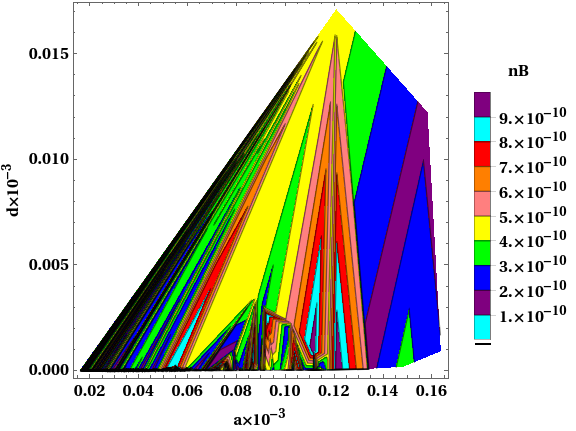}
		\includegraphics[width=0.4\textwidth]{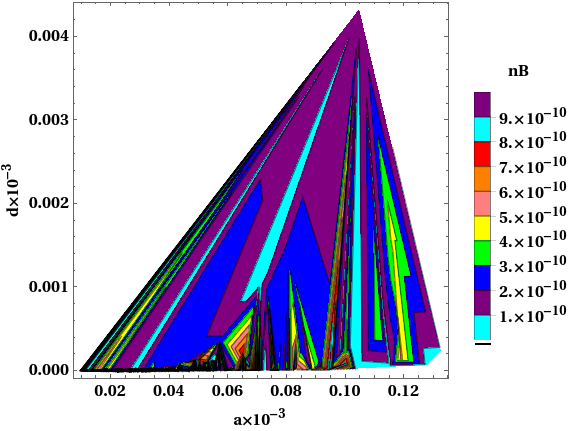}\\
		\caption{Contour plots relating the model parameter a and d(i.e. the perturbation) to the baryon asymmetry of the Universe. The left panel is for NH and right panel for IH.}\label{cont2}
	\end{center}
	\end{figure}

We choose the parameter space in such a way so as to fulfill the constraints coming from various phenomenologies. Considering the lightest RHN in TeV scale is a significant characteristic for vanilla leptogenesis in Scotogenic model\cite{Hugle:2018qbw,Mahanta:2019gfe}. A lower bound of about 10 TeV is set for $\rm N_{1}$, which has been verified in many literatures\cite{Borah:2018rca,Hugle:2018qbw}. Again, an inert Higgs doublet cannot possibly produce the observed relic density in the mass regime $\rm M_{W} < \rm M_{DM} \le 550$ GeV, also called the IHDM desert. Thus, we have considered the lightest of the inert scalar doublet in the range given in Table.\eqref{TAB1} in order to abide by the bounds from Planck limit to be a probable dark matter candidate and also to check its viability in the range 400-800 GeV. Again the charged scalar($\eta_{+}$) of the inert doublet is taken to be ($\eta_{R}^{0}+5$) GeV, following the constraints from LEP II\cite{Lundstrom:2008ai}. The choice of quartic coupling between the SM Higgs and inert doublet $\lambda_{5} \ne 0$ is to cause violation of the lepton number. \\

Now, we diagonalise the light neutrino mass matrix ($\rm M_{\nu l}$) by:
\begin{equation}
\rm U^{T}_{PMNS}\rm M_{\nu l}\rm U_{PMNS}= \tilde{\rm M_{\nu l}}= \begin{pmatrix}
m_{1}  & 0 & 0 \\
0 & m_{2} & 0\\
0 & 0 & m_{3}\\
\end{pmatrix}
\end{equation}
where,
\begin{equation}
\rm U_{\text{PMNS}}=\left(\begin{array}{ccc}
c_{12}c_{13}& s_{12}c_{13}& s_{13}e^{-i\delta}\\
-s_{12}c_{23}-c_{12}s_{23}s_{13}e^{i\delta}& c_{12}c_{23}-s_{12}s_{23}s_{13}e^{i\delta} & s_{23}c_{13} \\
s_{12}s_{23}-c_{12}c_{23}s_{13}e^{i\delta} & -c_{12}s_{23}-s_{12}c_{23}s_{13}e^{i\delta}& c_{23}c_{13}
\end{array}\right) U_{\text{Maj}}
\label{matrixPMNS}
\end{equation}
is the PMNS (Pontecorvo-Maki-Nakagawa-Sakata) matrix and $c_{ij} = \cos{\theta_{ij}}, \; s_{ij} = \sin{\theta_{ij}}$ and $\delta$ is the leptonic Dirac CP phase. The diagonal matrix $\rm U_{\text{Maj}}=\text{diag}(1, e^{i\alpha}, e^{i(\beta+\delta)})$  contains the Majorana CP phases $\alpha, \beta$.

The diagonal mass matrix of the light neutrinos can be written  as, $\tilde{\rm M_{\nu l}}
= \text{diag}(m_1, \sqrt{m^2_1+\Delta m_{21}^2}, \sqrt{m_1^2+\Delta m_{31}^2})$ for normal hierarchy and  $\tilde{\rm M_{\nu l}} = \text{diag}(\sqrt{m_3^2+\Delta m_{23}^2-\Delta m_{21}^2}, 
\sqrt{m_3^2+\Delta m_{23}^2}, m_3)$ for inverted hierarchy. We then numerical solve the model parameter, thereby generating the light neutrino mass matrix, the Yukawa coupling matrix and the neutrino mixing matrix. \\
A significant experimental technique of detecting neutrino mass is by the process of neutrinoless double beta decay ($0\nu\beta\beta$) \cite{Mohapatra:1986su,Giunti:2004vv,Barry:2013xxa,Borgohain:2017inp}. Some of the well known experiments related to it are KamLAND-Zen\cite{Kamland2,kamland}, GERDA\cite{gerda,GERDA2}, KATRIN\cite{katrin2,KATRIN}. We measure the effective neutrino mass $|\rm m_{\beta\beta}|$ expressed by the formula,
\begin{equation}
|\rm m_{\beta\beta}|= \sum_{k=1}^3\rm m_{k}\rm U_{ek}^{2}\label{eq:12}
\end{equation}
where, $\rm U_{ek}^{2}$ are the elements of the neutrino mixing matrix with $k$ holding up the generation index. This eq.\eqref{eq:12} can be further expressed as,
\begin{equation}
|\rm m_{\beta\beta}|= |\rm m_{1}\rm U_{ee}^{2} + \rm m_{2}\rm U_{e\nu}^{2} + \rm m_{3}\rm U_{e\tau}^{2}|.
\end{equation}
It is important to check if the model obeys the bound of the effective mass with the lightest neutrino mass so that we can relate the current light neutrino parameters giving correct hints to ongoing experiments and their future sensitivity. 

\begin{figure}[h]
	\begin{center}
		\includegraphics[width=0.4\textwidth]{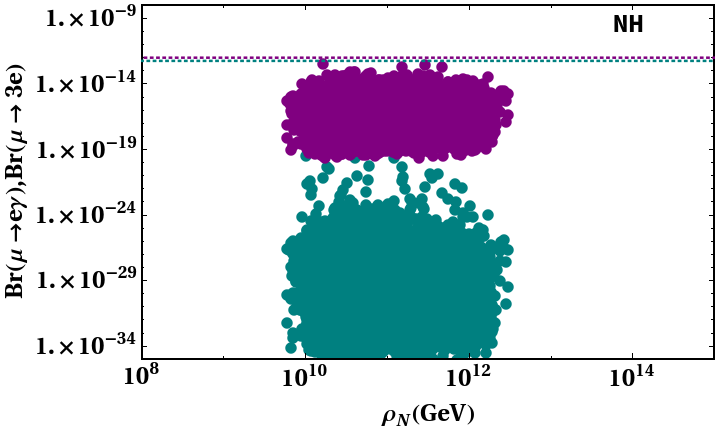}
		\includegraphics[width=0.4\textwidth]{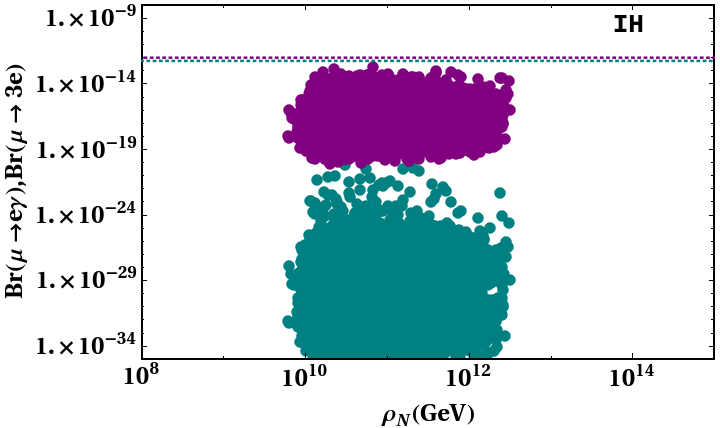}\\
		\caption{$Br(\mu\rightarrow e\gamma)$ and $Br(\mu\rightarrow 3e)$ as a function of $\rho_{N}$ (where $\rho_{N}=(\frac{M_{N}}{m_{\eta^{+}}})^{2}$) for NH and IH. The dashed horizontal lines are the recent upper bounds}.\label{br1}
	\end{center}
\end{figure}
\begin{figure}[h]
	\begin{center}
		\includegraphics[width=0.4\textwidth]{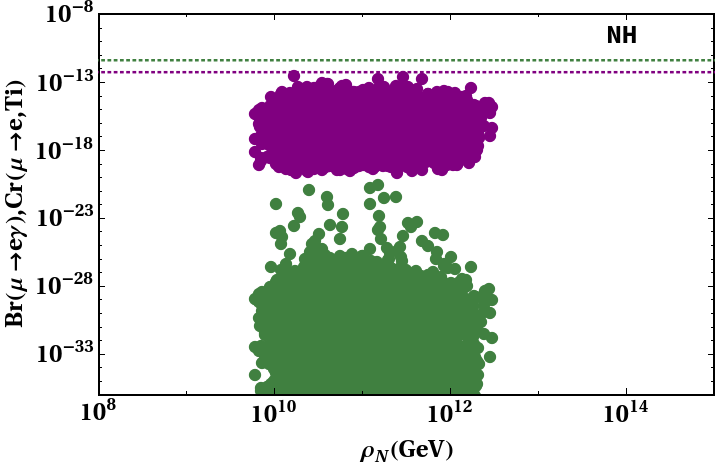}
		\includegraphics[width=0.4\textwidth]{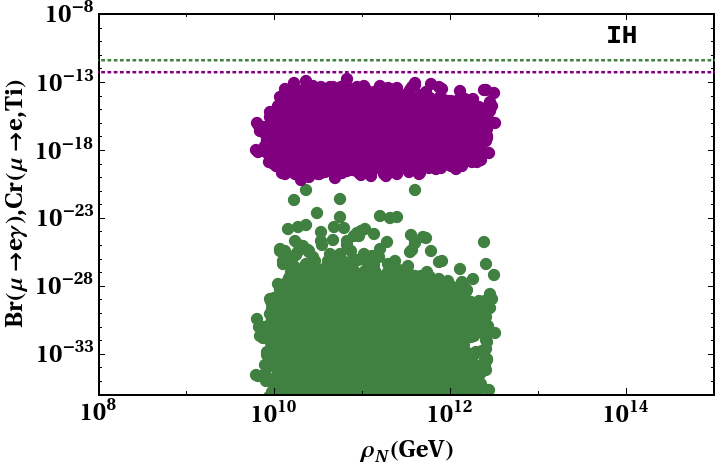}\\
		\caption{$Br(\mu\rightarrow e\gamma)$ and $Cr(\mu\rightarrow e,Ti)$ as a function of $\rho_{N}$ (where $\rho_{N}=(\frac{M_{N}}{m_{\eta^{+}}})^{2}$) for NH and IH. The dashed horizontal lines are the recent upper bounds.}\label{br2}
	\end{center}
\end{figure}

\begin{figure}[h]
	\begin{center}
		\includegraphics[width=0.4\textwidth]{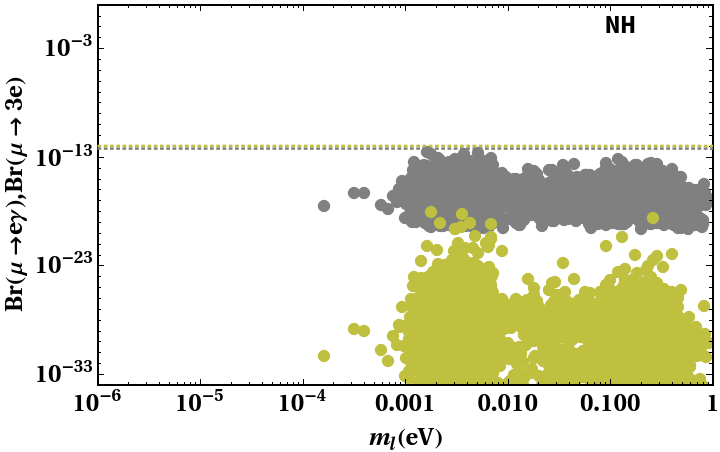}
		\includegraphics[width=0.4\textwidth]{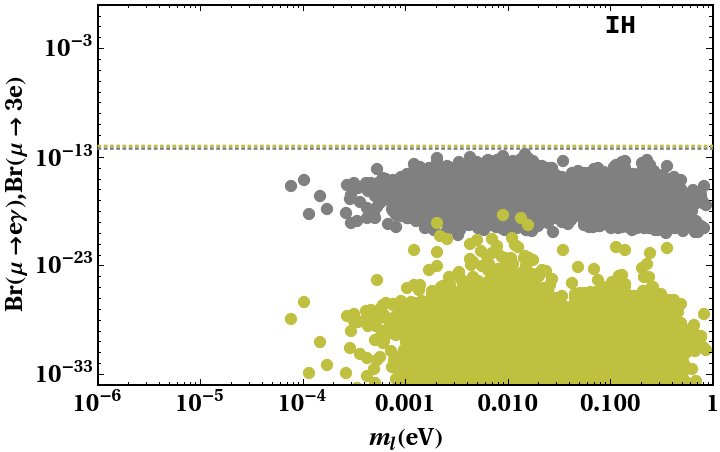}\\
		\caption{$Br(\mu\rightarrow e\gamma)$ and $Br(\mu\rightarrow 3e)$ as a function of lightest neutrino mass eigenvalue($m_{l}$) for NH and IH. The dashed horizontal lines are the recent upper bounds.}\label{br3}
	\end{center}
\end{figure}

\begin{figure}[h]
	\begin{center}
		\includegraphics[width=0.4\textwidth]{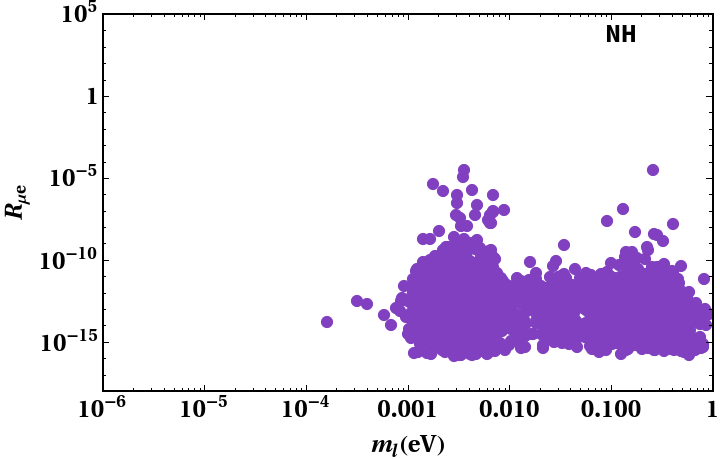}
		\includegraphics[width=0.4\textwidth]{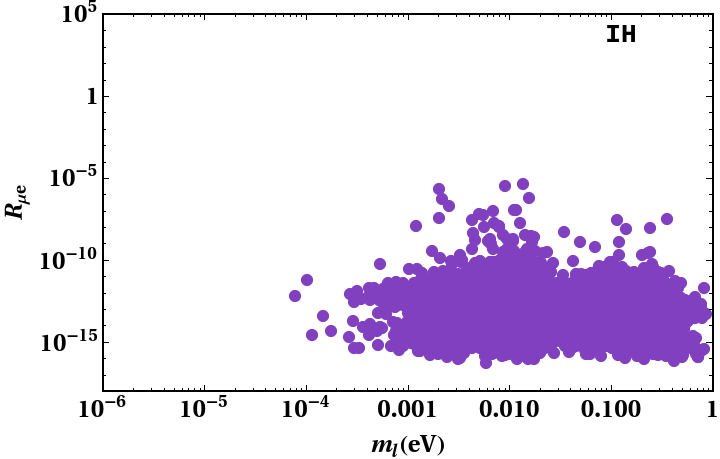}\\
		\caption{$R_{\mu e}$ as a function of lightest neutrino mass eigenvalue($m_{l}$) for NH and IH. Here, $R_{\mu e}= \frac{Br(\mu\rightarrow 3e)}{Br(\mu\rightarrow e\gamma)}$ .}\label{br4}
	\end{center}
\end{figure}

\begin{figure}[h]
	\begin{center}
		\includegraphics[width=0.4\textwidth]{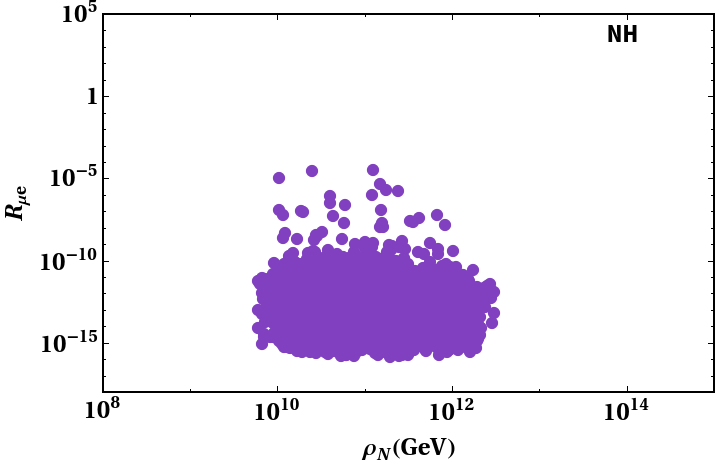}
		\includegraphics[width=0.4\textwidth]{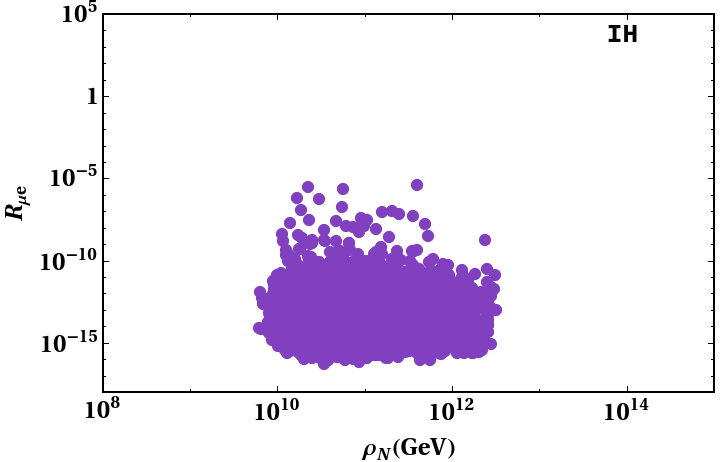}\\
		\caption{$R_{\mu e}$ as a function of $\rho_{N}$ for NH and IH. Here, $R_{\mu e}= \frac{Br(\mu\rightarrow 3e)}{Br(\mu\rightarrow e\gamma)}$.}\label{br5}
	\end{center}
\end{figure}
From Fig.\eqref{eff1}, we see that the effective mass of the light neutrinos are consistent with the KamLAND-Zen experiment for $\rm m_{l}= 10^{-4}- 10^{-2}$ eV incase of both NH and IH. A variation between mass of the lightest RHN($\rm M_{N_{1}}$) and BAU for NH/IH is depicted in the first row of Fig.\eqref{bau1}. We see that the parameter space taken for $\rm M_{N_{1}}$ is compatible for generating the observed BAU for both NH/IH. The second row of Fig.\eqref{bau1} shows that $\lambda_{5}= 10^{-6}- 10^{-4}$ has concentrated points satisfying the Planck limit for BAU for NH/IH. However, we have fewer points below $\lambda_{5}=10^{-6}$ which satisfy the bound for BAU. Thus, we can conclude that the space $\lambda_{5}= 10^{-6}- 10^{-4}$ is consistent with the BAU. Again from the variation plot between the lightest active neutrino mass eigenvalue($\rm m_{l}$) and BAU, we get a constraint region of $\rm m_{l}$ which satisfies the Planck limit for BAU both for NH/IH. Also, interestingly the constraint region of $\rm m_{l}$ falls within the bounds given by Planck for the summation of the light neutrino masses. The fourth row of Fig.\eqref{bau1} shows that the variation of CP violating phase($\delta$) with the baryon asymmetry of the Universe. Here, we observe that $\delta$ value above $10^{-2}$ satisfies the Planck limit for BAU for both NH/IH. A plot of BAU as a function of effective neutrino mass is evaluated as can be seen from Fig.\eqref{eff2}. It is thereby observed that the effective neutrino mass ranging from $10^{-2}-10^{-1}$ eV is successful in generating the correct BAU for NH, similar to that incase of IH. In Fig.\eqref{cont1}, we show a contour plot between the model parameters a and b with the effective mass of light neutrinos. And from Fig.\eqref{cont1}, we see a co relation between a and b with the baryon asymmetry of the Universe.  From these plots, we obtain the range of the model parameters, i.e $a\simeq 0.8\times 10^{-4}-0.13\times 10^{-4}$ and $b\simeq 0.2-0.4$ which further gives rise to a constraint in the Yukawa coupling matrix with its elements having values less than 3.We have considered a perturbation d in order to break the degeneracy of the RHN as well as to deviate from the $\mu-\tau$ symmetry. So, in Fig.\ref{cont2} we show a contour plot so as to constraint d w.r.t the baryon asymmetry of the Universe.\\
We have studied the LFV processes within the model. LFV obervables are ploted against different parameters defined within the parameter spaces of the model. Relevant formulae for LFV proceses $l_{\alpha}\rightarrow l_{\beta}\gamma$ , $l_{\alpha}\rightarrow 3l_{\beta}$ and $\mu-e$ conversion are discussed in section\eqref{s3}. We have defined the ratio of branching ratio of two LFV decay $l_{\alpha}\rightarrow l_{\beta}\gamma$ and $l_{\alpha}\rightarrow 3l_{\beta}$ as $R_{\mu e}$\cite{Takashi} i.e
\begin{equation}
\rm R_{\mu e}=\frac{Br(l_{\alpha}\rightarrow 3l_{\beta})}{Br(l_{\alpha}\rightarrow l_{\beta}\gamma)}
\end{equation}

In case of $\mu-e$ conversion the Z-penguins gives a very little contribution compared to $\gamma$-penguines. In this sutuation dipole operator will dominate the conversion rate. So, the coversion rate will have a very simple relation given below-
\begin{equation}
\rm \frac{CR(\mu-e,,Nucleus)}{Br(\mu\rightarrow e\gamma)}\approx \frac{f(Z,N)}{428}
\end{equation}

Where $\rm f(Z,N)$ is a funation which depends on the nucleus and ranges from 1.1 to 2.2 for nuclei of interest i.e titanium.

We have computed all the branching ratio of LFV decays and conversion ratio taking  consideration of contraints coming from the model. Variation of $\rm Br(\mu\rightarrow e\gamma)$ and $\rm Br(\mu\rightarrow 3e)$ as a function of $\rho_{N}$ (where $\rho_{N}=(\frac{M_{N}}{m_{\eta^{+}}})^{2}$) is depicted in Fig.\eqref{br1}. In this case for both the mass orderings, we get $\rm Br(\mu\rightarrow e\gamma)$ in the range $10^{-18}$ to $10^{-13}$ and $\rm Br(\mu\rightarrow 3e)$ spanning from $10^{-33}$ to $10^{-23}$, which are consistant with current and near future experimental limits. Similarly we have ploted the variation of $\mu-e$ conversion ratio in Fig.\eqref{br2} against  $\rho_{N}$ which is also in the experimental limits. Variation of lightest neutrino mass eigenvalue for both the mass orderings with both the branching ratio $\rm Br(\mu\rightarrow e\gamma)$ and $\rm Br(\mu\rightarrow 3e)$ is given in Fig.\eqref{br3}. From this we can see that for both the mass ordering results are consistant with experimental limit.

In Fig.\eqref{br4} we have ploted the variation of $R_{\mu e}$ which is the ratio of two LFV decays $l_{\alpha}\rightarrow l_{\beta}\gamma$ and $l_{\alpha}\rightarrow 3l_{\beta}$ against the lightest neutrino mass eigenvalue for both the mass orderings. From this we can infer that in case of both NH/IH ,$l_{\alpha}\rightarrow l_{\beta}\gamma$ decay supress the $l_{\alpha}\rightarrow 3l_{\beta}$ decay in our prameter spaces. Also, we can see that the parameter space of lightest active neutrino mass for which we get this kind of suppression is in the range $ 10^{-3}-1$eV. Further, we have also generated a plot (Fig.\eqref{br5}) depicting the co relation between $R_{\mu e}$ and $\rho_{N}$ and it is observed that the viable range for $\rho_{N}$ is $10^{10}-10^{12}$GeV. 
We have shown a common parameter space of the model parameters, satisfying the baryon asymmetry of the Universe in Table\ref{TAB3}.

\begin{table}[h]
	\begin{center}
		\begin{tabular}{|c|c|c|}
			
		\hline 
			Parameter & NH & IH \\ 
			\hline
			a & $0.8\times 10^{-4}-0.13\times 10^{-4}$ & $0.6\times 10^{-4}-0.9\times 10^{-4}$  \\
			\hline 
			b & $0.2-0.4$ & $0.1-0.16$ \\
			\hline  
				d & $10^{-6}-0.15\times 10^{-4}$ & $0.9\times 10^{-6}-10^{-6}$ \\
			\hline  		
		\end{tabular} 
	\end{center}
\caption 	{Model parameters of the model and their respective parameter space satisfying the Planck limit for baryon asymmetry of the Universe.} \label{TAB3}
\end{table}
We see from Table.\ref{TAB3} that a larger parameter space of a,b and d  are consistent with the Planck limit for BAU incase of NH as compared to that of IH.
\section{Conclusion}\label{s6}
In our work, we have basically realised Ernest Ma's Scotogenic model with the help of discrete symmetries $A_{4}\times Z_{4}$. A $\mu-\tau$ symmetric neutrino mass matrix is obtained, which is thereafter broken by introducing a perturbation to it. This mechanism is required for the generation of realistic neutrino mixing i.e. non zero $\theta_{13}$, deviation of $\theta_{23}$ from maximality and small correction in solar mixing angle $\theta_{12}$. As $r_{i}\propto \frac{1}{M_{N_{i}}}$, we have broken the degenracy in the masses of the RHN with the implementation of a perturbation which further breaks the $\mu-\tau$ symmetry. We have taken four free parameters, $ M_{N_{1}}, M_{N_{2}}, \eta_{R}^{0}$ and $\lambda_{5}$ whose values are mentioned above. By this choice of the parameter space, we have shown its consistency with various experimental and cosmological bounds. The lightest of the RHN, decays to produce lepton asymmetry which is further converted into BAU. Thus, the parameter space taken into account for the generation of the BAU are seen to follow the Planck limit for BAU. From the third row of Fig.\eqref{bau1}, we can see that the lightest active neutrino mass eigenvalue obtained from the model satisfies the Planck limit for BAU and consecutively obeys the Planck limit for the summation of light neutrino masses for NH/IH. Thus, the model is viable in connecting BAU and $0\nu\beta\beta$, also satisfying the bounds coming from neutrino oscillation data. We have also calculated the effective mass of the light neutrinos and shown the results in Fig.\eqref{eff1}. The lightest active neutrino mass eigenvalue for both NH/IH is seen to satisfy the KamLAND-Zen limit for effective mass of light neutrinos. Additional to this, a contour plot co-relating the model parameters with BAU is also studied. The conclusion we can draw from it is that the parameter space obtained from the model falls within the experimental bounds, thereby constraining the Yukawa coupling matrix. Furthermore, we have computed the branching ratios, $\rm Br(\mu\rightarrow e\gamma)$ and $\rm Br(\mu\rightarrow 3e)$ of LFV decays along with the $\mu-e$ conversion ratio. A variational plot between the branching ratios and $\rho_N$ is shown in Fig.\eqref{br1}, which is conistent with the current and the near future experimental bounds. Also, a similar plot for conversion ratio is also shown in Fig.\eqref{br2}, which also satisfies the current upper bound. We have also studied a co-relation plot between the branching ratios and the lightest neutrino mass eigenvalue and see that it also obeys the experimental upper bounds for the branching ratios. It is predicted well from Fig.\eqref{br4} and Fig.\eqref{br5}  that $l_{\alpha}\rightarrow l_{\beta}\gamma$ decay supress the $l_{\alpha}\rightarrow 3l_{\beta}$ decay incase of both NH/IH. We can say that the results hardly show any change depending on the mass hierarchies. Overall, this realisation of the Scotogenic model by discrete symmetries, with the considerations on the free parameters taken from various bounds is viable for studying neutrino as well as cosmological phenomenologies. 

\section{APPENDIX:}
\subsection{$A_{4}$ product rules:}
$A_4$ is discrete group of even permutation of four objects. It has three inequivalent onedimensional representation 1, $1^{\prime}$ and$1^{\prime\prime}$ a irreducible three dimensional representation 3. Product of the singlets and triplets are given by-

\begin{equation}
1 \otimes 1=1 \nonumber
\end{equation} 
\begin{equation}
1^{\prime}\otimes 1^{\prime}=1^{\prime\prime} \nonumber
\end{equation}
\begin{equation}
1^{\prime}\otimes1^{\prime\prime}=1 \nonumber
\end{equation}
\begin{equation}
1^{\prime\prime} \otimes 1^{\prime\prime}=1^{\prime} \nonumber
\end{equation}
\begin{equation}
3\otimes3=1\oplus1^{\prime}\oplus1^{\prime\prime}\oplus 3_{A}\oplus 3_{S}
\end{equation}

where subscripts A and S stands for “asymmetric” and “symmetric” respectively. If we
have two triplets ($a_1$, $a_2$, $a_3$) and ($b_1$, $b_2$, $b_3$), their products are given by

\begin{equation}
1 \approx a_1b_1 + a_2b_3 + a_3b_2 \nonumber
\end{equation}
\begin{equation}
1^\prime \approx a_3b_3 + a_1b_2 + a_2b_1 \nonumber
\end{equation}
\begin{equation}
1^{\prime\prime} \approx a_2b_2 + a_3b_1 + a_1b_3 \nonumber
\end{equation}
\begin{equation}
3_S \approx \left(\begin{array}{c}
2a_{1}b_{1}-a_{2}b_{3}-a_3b_2\\
2a_{3}b_{3}-a_{1}b_{2}-a_2b_1\\
2a_{2}b_{2}-a_{1}b_{3}-a_3b_1 \end{array}\right)\nonumber
\end{equation}
\begin{equation}
3_A \approx \left(\begin{array}{c}
a_{2}b_{3}-a_{3}b_{2}\\
a_{1}b_{2}-a_2b_1\\
a_{3}b_{1}-a_{1}b_3 \end{array}\right)\nonumber
\end{equation}

\subsection{Loop functions:}
We present in this appendix the loop functions relevant for the computation of the LFV
observables. These are

\begin{equation}
F_{2}(x)=\frac{1-6x+3x^2+2x^3+6x^2\log(x)}{6-(1-x)^{4}}
\end{equation}
\begin{equation}
G_{2}(x)=\frac{2-9x+18x^2-11x^3+6x^3\log(x)}{6-(1-x)^{4}}
\end{equation}

\begin{equation}
D_{1}(x,y)=-\frac{1}{(1-x)(1-y)}-\frac{x^2 \log(x)}{(1-x)^{2}(x-y)}-\frac{y^2\log(y)}{(1-y)^{2}(y-x)}
\end{equation}

\begin{equation}
D_{2}(x,y)=-\frac{1}{(1-x)(1-y)}-\frac{x\log(x)}{(1-x)^{2}(x-y)}-\frac{y \log(y)}{(1-y)^{2}(y-x)}
\end{equation}
\subsection{VEV ALIGNMENT OF THE FLAVONS}
	In this section we will evaluate the vev alignment of the flavons considered in the model by minimizing the potential and solving it simultaneously. The relevent potential can be written in terms of the field $\chi_{T}$, $\chi_{S}$ and their interaction terms. The interaction terms are forbidden in the model due the additional $Z_{4}$ symmetry. The potential is given by-
	
	\begin{equation}
		V=V(\chi_{T})+V(\chi_{S})+V_{int}
		\end{equation}
		with,
		\begin{equation}
		V(\chi_{T})=-m^{2}_{\chi_{T}}(\chi^{\dagger}_{T}\chi_{T})+\lambda_{1}(\chi^{\dagger}_{T}\chi_{T})^{2}
		\end{equation}
		and
		\begin{equation}
		V(\chi_{S})=-m^{2}_{\chi_{S}}(\chi^{\dagger}_{S}\chi_{S})+\lambda_{1}(\chi^{\dagger}_{S}\chi_{S})^{2}
		\end{equation}
		The triplet flavons can be written in the form-
		\begin{equation}
		<\chi_{T}>=(\chi_{T_{1}},\chi_{T_{2}},\chi_{T_{3}}),
		\\
		<\chi_{S}>=(\chi_{S_{1}},\chi_{S_{2}},\chi_{S_{3}})
		\end{equation}
		Considering $A_{4}$ product rule the potential terms for $\chi_{T}$ will be of the form -
		\begin{equation}
		\begin{split}
		V(\chi_{T})& =-\mu^{2}(\chi^{\dagger}_{T_{1}}\chi_{T_{1}}+\chi^{\dagger}_{T_{2}}\chi_{T_{3}}+\chi^{\dagger}_{T_{3}}\chi_{T_{2}})\\ & +\lambda_{2}[(\chi^{\dagger}_{T_{1}}\chi_{T_{1}}+\chi^{\dagger}_{T_{2}}\chi_{T_{3}}+\chi^{\dagger}_{T_{3}}\chi_{T_{2}})^{2}+(\chi^{\dagger}_{T_{3}}\chi_{T_{3}}+\chi^{\dagger}_{T_{2}}\chi_{T_{1}}+\chi^{\dagger}_{T_{1}}\chi_{T_{2}})\\  &\times(\chi^{\dagger}_{T_{2}}\chi_{T_{2}}+\chi^{\dagger}_{T_{1}}\chi_{T_{3}}+\chi^{\dagger}_{T_{3}}\chi_{T_{1}})+(2\chi^{\dagger}_{T_{1}}\chi_{T_{1}}-\chi^{\dagger}_{T_{2}}\chi_{T_{3}}-\chi^{\dagger}_{T_{3}}\chi_{T_{2}})^{2}\\ &+2(2\chi^{\dagger}_{T_{3}}\chi_{T_{3}}-\chi^{\dagger}_{T_{1}}\chi_{T_{2}}-\chi^{\dagger}_{T_{2}}\chi_{T_{1}})\times(2\chi^{\dagger}_{T_{2}}\chi_{T_{2}}-\chi^{\dagger}_{T_{3}}\chi_{T_{1}}-\chi^{\dagger}_{T_{1}}\chi_{T_{3}})]
		\end{split}
		\end{equation}
	
	Taking derivative with respect to $\chi_{T_{1}}$, $\chi_{T_{2}}$ and  $\chi_{T_{3}}$ and equating it to zero will give the minimization condition. After solving we will get three set of solutions which are basically the possible alignments given by-
	
	\begin{equation}
		\begin{split}
		(1)\hspace{0.2cm} & \chi_{T_{1}} \rightarrow \frac{\mu_2{}}{\sqrt{10\lambda_{2}}},\chi_{T_{2}} \rightarrow 0,\chi_{T_{3}} \rightarrow 0 \implies <\chi_{T}>=\frac{\mu_2}{\sqrt{10\lambda_{2}}}(1,0,0)\\
		(2)\hspace{0.2cm} & \chi_{T_{1}} \rightarrow \frac{\mu_2{}}{2\sqrt{3\lambda_{2}}},\chi_{T_{2}} \rightarrow \frac{\mu_2{}}{2\sqrt{3\lambda_{2}}},\chi_{T_{3}} \rightarrow \frac{\mu_2{}}{2\sqrt{3\lambda_{2}}} \implies <\chi_{T}>=\frac{\mu_2}{2\sqrt{3\lambda_{2}}}(1,1,1)\\
		(3)\hspace{0.2cm} & \chi_{T_{1}} \rightarrow \frac{2\mu_2{}}{\sqrt{51\lambda_{2}}},\chi_{T_{2}} \rightarrow -\frac{\mu_2{}}{\sqrt{51\lambda_{2}}},\chi_{T_{3}} \rightarrow -\frac{\mu_2{}}{\sqrt{51\lambda_{2}}} \implies <\chi_{T}>=\frac{\mu_2}{\sqrt{51\lambda_{2}}}(2,-1,-1)
		\end{split}
		\end{equation}
	Similarly we get the solutions for $\chi_{S}$. We have used the first set of solution to generate the charge lepton mass matrix and second set of solution to get the Dirac neutrino mass.

\bibliographystyle{utphys}
\bibliography{ref}
\end{document}